\documentclass[prl,aps,twocolumn,showpacs,superscriptaddress,longbibliography]{revtex4-2}

\usepackage{ulem}
\usepackage{graphicx}
\usepackage{amsfonts}
\usepackage{epstopdf}
\usepackage{color}
\usepackage[dvipsnames,svgnames]{xcolor} 

\begin{document}

\title{Multiple symmetry breaking induced by weak damping in the Fermi-Pasta-Ulam-Tsingou recurrence process}

\author{Guillaume Vanderhaegen}
\affiliation{University of Lille, CNRS, UMR 8523-PhLAM-Physique des Lasers Atomes et Mol\'ecules, F-59000 Lille, France}
\author{Pascal Szriftgiser}
\affiliation{University of Lille, CNRS, UMR 8523-PhLAM-Physique des Lasers Atomes et Mol\'ecules, F-59000 Lille, France}
\author{Alexandre Kudlinski}
\affiliation{University of Lille, CNRS, UMR 8523-PhLAM-Physique des Lasers Atomes et Mol\'ecules, F-59000 Lille, France}
\author{Andrea Armaroli}
\affiliation{University of Lille, CNRS, UMR 8523-PhLAM-Physique des Lasers Atomes et Mol\'ecules, F-59000 Lille, France}
\author{Matteo Conforti}
\affiliation{University of Lille, CNRS, UMR 8523-PhLAM-Physique des Lasers Atomes et Mol\'ecules, F-59000 Lille, France}
\author{Arnaud Mussot}
\affiliation{University of Lille, CNRS, UMR 8523-PhLAM-Physique des Lasers Atomes et Mol\'ecules, F-59000 Lille, France}
\author{Stefano Trillo} 
\affiliation{Department of Engineering, University of Ferrara, 44122 Ferrara, Italy}


\begin{abstract}
We show that even small dissipation can strongly affect the Fermi-Pasta-Ulam-Tsingou recurrence phenomenon. Taking the fully nonlinear stage of modulational instability as an experimentally accessible example, we show that the linear attenuation induces the symmetry of the recurrence to be broken through separatrix crossing occurring at multiple critical values of the attenuation. We provide experimental evidence for this phenomenon in a fiber optics experiment designed in such a way that the effective losses can be carefully tailored by techniques based on Raman amplification.
\end{abstract} 

\date{\today} 
\maketitle
{\it Introduction}. The celebrated phenomenon of recurrence named after Fermi-Pasta-Ulam-Tsingou (FPUT) \cite{FPUToriginal} stands for the ability of a nonlinear multimodal Hamiltonian system to repeatedly recover its initial condition after the energy has spread from the excited mode to all the others, in contrast to the principle of statistical equipartition. After almost 70 years \cite{Ford1992,Berman2005}, the investigation of this phenomenon is still very active, inspiring new results and opening to new realms (e.g., random nonlinear waves) \cite{Trillo2016,Guasoni2017,Lvov2018,Pace2019b,Zaleski2020,Dematteis2020}. Yet, understanding the observed FPUT dynamics in real nonlinear systems requires to analyze the impact of ingredients which are usually neglected.
Damping, which is often unavoidable in physical systems, is likely the most important one, especially in the perspective of characterizing the long-term thermalization process \cite{Pace2019b,Wabnitz2014}.

In this Letter, we show how the FPUT scenario is profoundly altered by (even weak) linear damping. FPUT is induced via the ubiquitous phenomenon of modulational instability (MI, the growth of slow periodic perturbations on top of a background/pump \cite{ZO2009,BF1967,Taniuti1968,Tai1986,Everitt2017,Leykam2021}) and ruled by the universal nonlinear Schr\"odinger equation (NLSE).
The nonlinear stage of MI has attracted a great deal of attention as a triggering mechanism of rogue waves \cite{Dudley2014,Onorato2013}. It turned out to be an incredibly rich and multi-faceted phenomenon whose understanding is slowly progressing thanks to recent efforts that range from theory 
\cite{ZG2013,Biondini2016,SotoCrespo2016,Armaroli_2017,Armaroli_2018,GS2018a,GS2018b,Biondini2018,Conforti2018,Trillo2019,Conforti2020} to experiments mainly concentrated in the areas of 
nonlinear optics \cite{VanSimaeys2001,Erkintalo2011,Mussot2014,Kibler2015,Mussot2018,Pierangeli2018,Kraych2019a,Kraych2019b,Goossens2019,Schiek2019,Nielsen2021,Vanderhaegen2021PNAS}
and hydrodynamics \cite{Chabchoub2011,Kibler2015,Kimmoun2016,Eeltink2020,Bonnefoy2020,Vanderhaegen2021PNAS}.
In this context, FPUT recurrences emerge, when MI is seeded by periodic perturbations (sideband pairs), as cycles of conversion and back-conversion between the sidebands and the pump. 
Importantly, two qualitatively different types of recurrence cycles coexist, due to the symmetry breaking nature of the MI, as demonstrated by recent experiments in optics \cite{Mussot2018,Pierangeli2018,Goossens2019}. 
The recurrence type is selected solely by the initial condition, and cross-over into the other type, under lossless conditions, is prevented by the conservation of the Hamiltonian.
Our aim is to show that linear damping causes, instead, the recurrences to undergo a dynamical change from one type to the other, through a phenomenon of loss-induced {\it separatrix crossing}. 
Remarkably, we find by perturbative arguments (based either on a Fourier mode-truncation or finite-gap theory  \cite{Coppini2020,Schober2021}) that such crossing occurs around multiple critical values of the linear damping coefficient.
We demonstrate this by reporting evidence for the first two critical loss values in a fiber optics experiment, where we reconstruct, via non-destructive measurements, the nonlinear evolution of MI in power and phase,
as the effective losses are accurately tailored via Raman amplification techniques \cite{Agrawalbook,AniaCastanon2006, Naveau2021}. Our results give evidence that weak losses qualitatively alter the FPUT scenario without suppressing MI \cite{Segur2007},
and establish quantitatively how to understand the impact of damping on FPUT experiments carried out in other areas, e.g. hydrodynamics \cite{Kimmoun2016}.

{\it Theory of separatrix crossing}. We start from the NLSE with damping written in dimensionless form
\begin{eqnarray}  \label{NLS}
 i\frac{\partial \psi}{\partial z} + \frac{1}{2} \frac{\partial^2 \psi}{\partial t^2} + |\psi|^2 \psi  = -i \frac{\alpha}{2} \psi,
\end{eqnarray}
where $\psi=E/\sqrt{P_0}$, $z=Z/Z_{nl}$, $t=(T-Z/V_g)/T_0$, are the normalized field, distance and retarded time (capital letters stand for real-world quantities), in units of total input power $P_0$, and associated nonlinear length $Z_{nl}=1/(\gamma P_0)$ and time $T_0=\sqrt{|\beta_2|Z_{nl}}$, respectively. Here $\gamma$ is the fiber nonlinear coefficient, $\beta_2$ is the dispersion, and $V_g$ the group-velocity. The key parameter is the normalized attenuation coefficient $\alpha = \alpha_P Z_{nl}$(km)$/4.34$ where $\alpha_P$ (dB/km) is the physical attenuation. 

\begin{figure}[ht]
\includegraphics[width=\columnwidth]{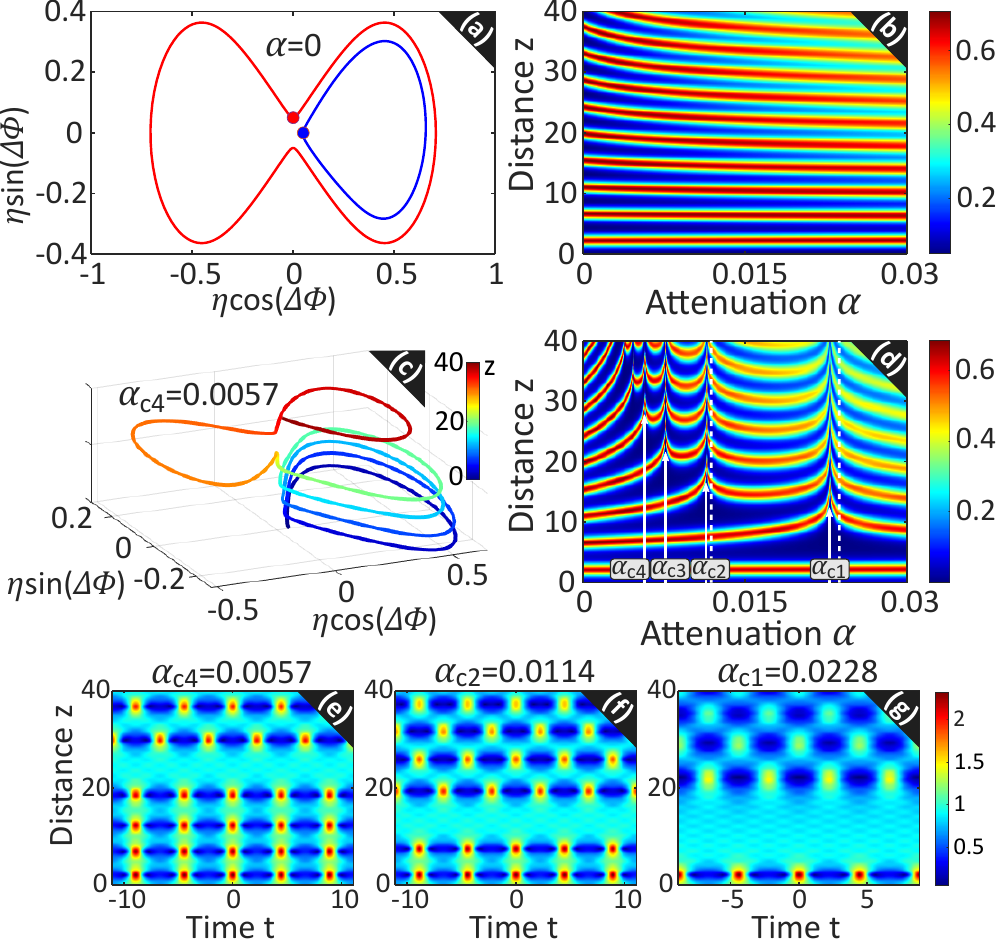}
\caption{(a) NLSE unperturbed ($\alpha=0$) evolutions of unshifted (blue) and shifted (red) FPUT type, projected over the phase-plane $(\eta \cos (\Delta \phi ), \eta \sin (\Delta \phi))$ (dots are initial conditions); 
(b, d) False color plot of sideband power fraction $\eta$ in the plane ($\alpha,z$) for initial values on the shifted (b) or unshifted (d) orbit.
The arrows in (d) indicate critical losses $\alpha_{cn}$ around which separatrix crossing occurs; vertical dashed lines are estimates of $\alpha_{c1,c2}$ from Eq. (\ref{alphac_IST}).
(c) Typical separatrix crossing in 3D space $(\eta \cos ( \Delta \phi), \eta \sin ( \Delta \phi), z)$ at $\alpha_{c4}$, cf. case (e). (e,f,g) Evolutions of $|\psi(z,t)|$ at sampled values $\alpha_{c1,c2,c4}$.
In all plots $\omega_0=\sqrt{2}$ (peak gain), and the injected sideband power fraction is 5\% ($\eta_0=0.05$).
}
\label{f1}
\end{figure}

We consider FPUT arising from induced MI which is ruled by Eq.~(\ref{NLS}) subject to the initial condition 
\begin{equation} 
\psi_0=\sqrt{p} \left[1+ \frac{a}{\sqrt{2}} e^{-i \phi_0} \left(e^{i \omega_0 t} +  e^{-i \omega_0 t} \right) \right],
\end{equation}
where, without loss of generality, we take a $\sqrt p$ amplitude pump modulated by symmetric sidebands with normalized input pulsation $\omega_0=2\pi f_{mod} \sqrt{|\beta_2|/(\gamma P_0)}$ (in the range $1<\omega_0<2\sqrt{p}$ to have only one unstable pair \cite{Erkintalo2011}), relative power $a^2 \ll 1$, and phase $\phi_0$, with $p \equiv (1+a^2)^{-1}$ to have $\int |\psi_0|^2 dt=1$.
In the unperturbed case ($\alpha=0$), two different FPUT types of recurrence coexist, as conveniently illustrated in Fig. \ref{f1}(a) by their projections on the phase-plane $(\eta \cos \Delta \phi, \eta \sin \Delta \phi)$,
where $\eta(z)$ is the first-order sideband power fraction and $\Delta \phi(z)=\phi_P-\phi_S$ is the relative phase between the pump and the sidebands \cite{Mussot2018}. 
We refer to the two types of FPUT orbits as unshifted (single loop, blue curve) or shifted (double loop, red curve) recurrence, since in the latter case, owing to the free running phase, two consecutive recurrences exhibit a shift of $\pi$ (half period in time) \cite{Mussot2018}.

In the unperturbed case the two orbits are separated by a separatrix which is known as Akhmediev breather \cite{Akhmediev1986,Akhmediev1987} and never cross each other. They are typically generated by inputs corresponding to the dots in Fig. \ref{f1}(a) representing weak modulations of the pump (i.e. the saddle point in the origin) with constant sideband fraction $\eta(z=0 )=\eta_0=\frac{a^2}{1+a^2}$, but different phase, $\Delta \phi_0=0$ (amplitude modulation) or $\Delta \phi_0=\pm \pi/2$ (frequency modulation), respectively.
A convenient way to assess the impact of the losses is to portray, as shown in Fig. \ref{f1}(b,d), the evolution of the sideband fraction $\eta(z)$, as obtained from numerical integration of the damped NLSE, as a function of the attenuation $\alpha$.
The periodicity in the vertical ($z$) direction accounts for the FPUT recurrence. 
When the shifted orbit is excited, the net effect of damping is smooth, inducing only a slight reduction of the FPUT period as evident from Fig.~\ref{f1}(b). Conversely, an initially unshifted orbit results in the complex scenario displayed in Fig. \ref{f1}(d). It is characterized by a succession of critical attenuation values $\alpha_{cn}$, $n=1,2,3,\ldots$ (with $\alpha_{c1} > \alpha_{c2} > \alpha_{c3} \ldots$), around which the FPUT recurrence slow down dramatically giving rise to extremely large (but still finite) recurrence distances. The underlying mechanism is the dynamical {\it separatrix crossing} from unshifted to shifted orbits which takes place, for any loss coefficient $\alpha_{cn} \le \alpha <  \alpha_{c(n-1)}$, at the closest passage to the initial condition after $n$ unshifted recurrences are completed. Conversely, for $\alpha_{c(n+1)} \le \alpha <  \alpha_{cn}$ also the $(n+1)$-th recurrence will be still unshifted and crossing occurs at the $(n+2)$-th recurrence.
This is conveniently illustrated, for $\alpha = \alpha_{c4}$, by the phase-space projection in Fig. \ref{f1}(c), where four unshifted revolutions are clearly visible along with the slowing down in the fourth passage close to the origin which precedes the crossing. For better clarity,  Fig. \ref{f1}(e-g) further show examples of spatio-temporal evolutions obtained at the critical values $\alpha_{c4}$ (same as in Fig. \ref{f1}(c)) and $\alpha_{c2,c1}$, showing indeed the appearance of the phase shift after $n=4$ or $n=2,1$ recurrences, respectively. 
It must be noticed also that the growth and decay cycles do not proceed at constant rate but slow down considerably (as in any nonlinear oscillator) when approaching the saddle point (i.e. at the return to the initial condition).
This effect is strongly enhanced at the critical values $\alpha_{cn}$, which is such that the evolution becomes locally iso-energetic (i.e., same local Hamiltonian) to the saddle in its closest passage after $n$ unshifted orbits.
This argument allows us to give, in the framework of a generalized three-mode approximation \cite{Mussot2018,Trillo1991a,SM}, the critical values $\alpha_{cn}$ as the implicit solutions of the following integral equation
\begin{equation} \label{critical}
H_0 = \frac{\alpha_{cn} \omega_0^2}{2}~\int_{0}^{n z_\mathrm{per}} e^{\alpha_{cn} z}~\eta(z) dz,
\end{equation}
where $H_0=\eta_0 (1-\eta_0) \cos 2 \Delta \phi_0 + (1-\omega_0^2/2) \eta_0 -\frac{3}{4} \eta_0^2$ is the input three-wave Hamiltonian \cite{Mussot2018}, and $z_\mathrm{per}$ is the period of the unperturbed unshifted orbit.
Equation (\ref{critical}) gives a quantitatively accurate estimate of the critical losses in the context of the three-wave approach. This thus validates the physical picture discussed above 
but shows discrepancy with the values obtained from the NLSE. This is due to the fact it underestimates the evolution period $z_\mathrm{per}$, which is a critical parameter \cite{SM}.

An alternative, more accurate, estimate of the values $\alpha_{cn}$ can be obtained by exploiting the perturbation theory based on the finite gap formulation of the inverse scattering transform for the NLSE \cite{Coppini2020}, which yields a particularly simple formula (details in \cite{SM}) 
\begin{equation} \label{alphac_IST}
\alpha_{cn}=  \frac{p^2 e_+ e_-}{g}  \frac{a^2}{n}, \;\; n=1,2,3,\ldots.
\end{equation}
where $e_+$ ($e_-$) are the growing (decaying) eigenvector of MI \cite{GS2018a,Trillo2019}, associated with the MI gain $g=g(\omega_0)=\omega_0\sqrt{4p-\omega_0^2}$.
Equation (\ref{alphac_IST}) turns out to be extremely accurate in the limit of small sidebands (i.e., $a=O(\epsilon)$, $\epsilon \ll 1$, which gives $\alpha=O(\epsilon^2)$, see \cite{SM}), 
but gives a reasonable approximation also for relatively large sidebands (5\%). This is depicted in Fig. \ref{f1}(d) by the vertical dashed lines corresponding to $\alpha_{c1,c2}$.

{\it Experiment}.
In order to demonstrate the existence of critical values of damping around which FPUT recurrence is qualitatively affected by the separatrix crossing phenomenon, we exploit a full fiber-optics setup, similar to the one in \cite{Mussot2018, Vanderhaegen2020OE, Vanderhaegen2020OL, Naveau2021, Vanderhaegen2021PNAS} and detailed in the Supplementary.

\begin{figure}[ht]
\includegraphics[width=8.6cm]{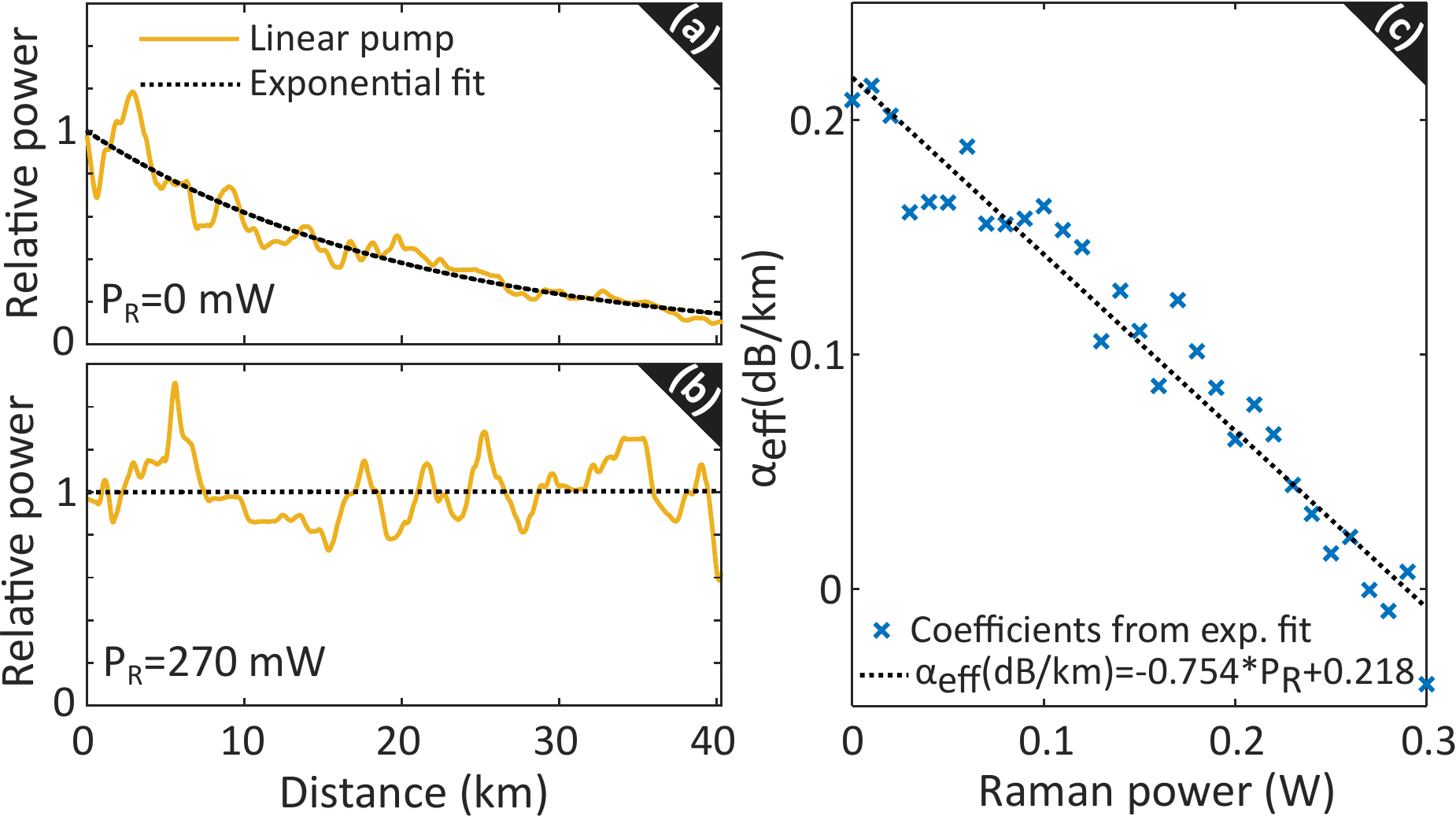}
\caption{(a,b) Power evolution of the backscattered signal, when the Raman pump is: (a) switched off; (b) optimized to get a nearly perfect loss compensation. (c) Effective loss $\alpha_\mathrm{eff}$ versus Raman pump power $P_R$.}
\label{fit}
\end{figure}

A key feature of the setup is the tailoring of the effective loss. To this end, we implemented a counter-propagating Raman pump to actively control the damping by tuning the Raman pump power. In Refs. \cite{Mussot2018, Vanderhaegen2020OE, Vanderhaegen2020OL, Naveau2021, Vanderhaegen2021PNAS}, we adjusted the Raman pump power to get an almost fully transparent optical fiber. By neglecting the Raman pump dissipation effect, a simplistic description of the loss compensation process allows us to assume that the Raman amplification term $g_R P_R^{Opt}$ is equal to the fiber loss $\alpha$ \cite{Agrawalbook}. Hence, the optical signal experiences a vanishing effective loss $\alpha_\mathrm{eff} \simeq 0$ dB/km during its propagation through the fiber. The compensation is almost perfect because the Raman amplifier always operates in a linear regime because signals to amplify are very short \cite{Vanderhaegen2021b}. By varying the Raman pump power from $0$ mW to $P_R^{Opt}$, we have been able to tune the effective loss $\alpha_\mathrm{eff}$ in a regime of extremely weak damping ranging from the intrinsic fiber loss value ($0.2$ dB/km) to $0$ dB/km, as illustrated in Fig. \ref{fit}.
For each value of Raman pump, we determine the effective damping as follows.
Weak square pulses of $48$ mW peak power are launched along the fiber. The nonlinear length $Z_{nl}$ being around $16$ km, we can assume that nonlinear effects are negligible. The HOTDR system allows to monitor their power evolution along the fiber length. As the pulses are mainly affected by dissipation, they experience an exponential decay, as illustrated in Fig.~\ref{fit}(a) with the Raman pump switched off.
Note that, in Fig.~\ref{fit}(a), the data cover twice the fiber length since we detect the backscattered signal (the generic distance is travelled forward and then backward before being detected).
By fitting the data with an exponential curve (dotted line in Fig.~\ref{fit}(a)), we extrapolate an intrinsic damping coefficient $\alpha_P\simeq 0.2084$ dB/km, in excellent agreement with the data-sheet value of $0.2$ dB/km.
The opposite limit that yields nearly perfect loss compensation for $P_R=P_R^{Opt}=270$ mW is shown in Fig.~\ref{fit}(b), which shows quasi-flat power profile.
We repeat these measurements for Raman pump powers in the range $0-300$ mW, with the upper value of $300$ mW yielding slight overcompensation (weak gain). The results are summarized in Fig.~\ref{fit}(c), from which we conclude that a nearly linear control of the loss coefficient can be achieved by tuning the Raman pump power. We make use of the linear fit in Fig.~\ref{fit}(c) to calibrate the effective damping in the successive experiments described below.

\begin{figure*}[ht]
\centering
\includegraphics[scale=0.9]{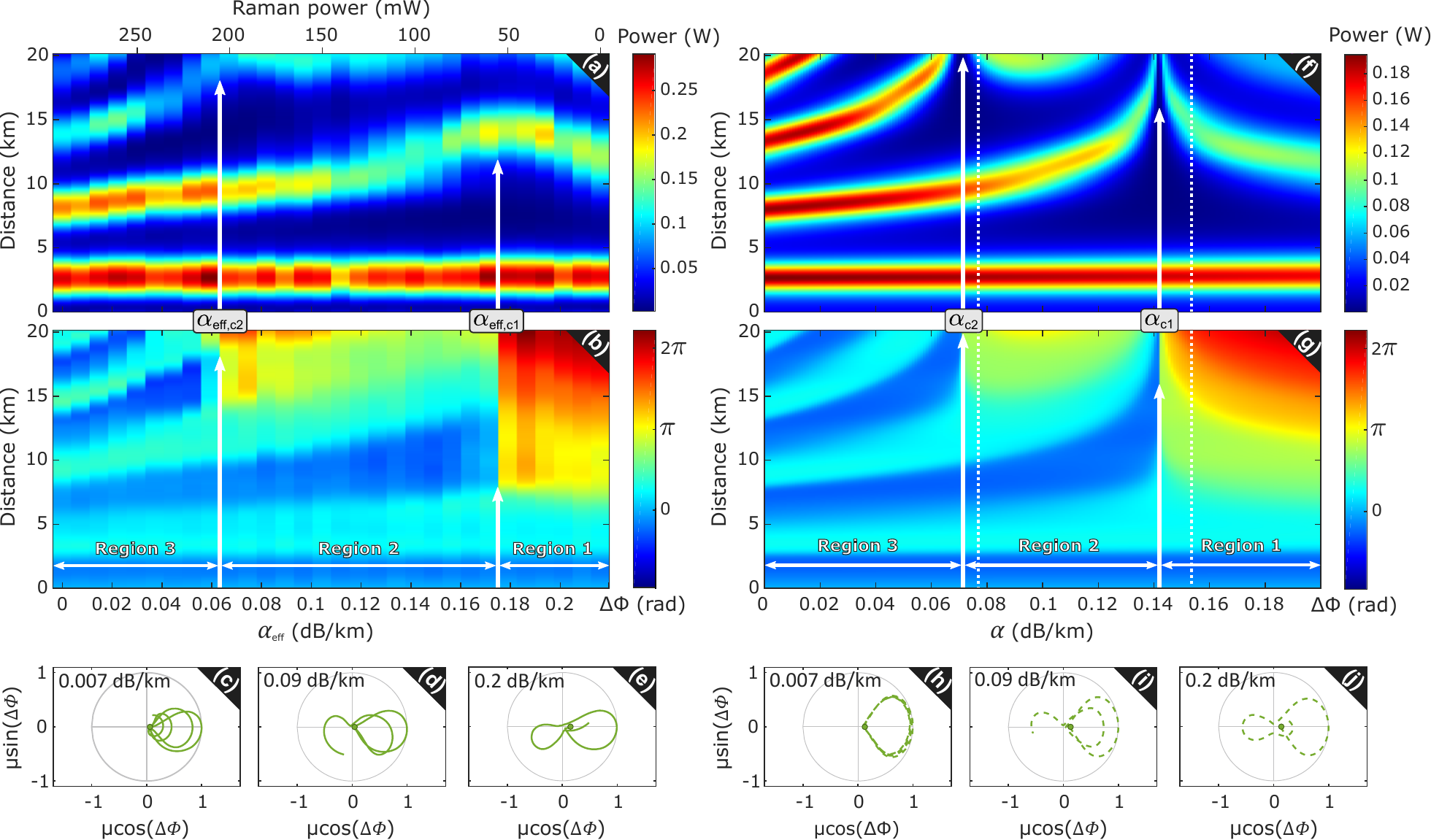}
\caption{Left column panels: Experimental data showing false color plots of signal power (a) and relative phase $\Delta \phi=\phi_P-\phi_S$ (b) in the plane $(\alpha_\mathrm{eff},Z)$ of effective loss and fiber distance; (c,d,e) projections of the measured evolutions in the phase-plane $(\mu \cos (\Delta \phi), \mu \sin (\Delta \phi))$ for values $\alpha_\mathrm{eff}(dB/km) = 0.007(c),0.09(d),0.2 (e)$ sampled in regions (1,2,3), respectively. $\mu$ stands for the signal power, normalized to the maximum value.
Right column panels (f-j) report the numerical counterpart of results in (a-e), obtained from integration of the damped NLSE [Eq. (1) with $\alpha=\alpha_\mathrm{eff}$]; white dotted lines in (f) stand for estimates of $\alpha_{c1,c2}$ predicted by finite-gap perturbation theory \cite{Coppini2020,SM}.}
\label{exp_vs_num_Ps_new}
\end{figure*}

{\it Experimental results.} We have designed the experimental setup in such a way that several recurrences can be observed along the fiber length in order to highlight the impact of weak effective fiber loss on the FPUT process. We have measured the evolution along the fiber of the signal (sideband) power and relative phase, for different values of effective loss, keeping the initial pump-signal relative phase to $\Delta \Phi(0)=0$ (unshifted orbits in the lossless case). The results showing the signal power evolution are summarized in Fig.~\ref{exp_vs_num_Ps_new}(a). Around optimal compensation ($\alpha_\mathrm{eff}=0$ achieved at $P_R \simeq 270$ mW), four nearly complete recurrence cycles are observed (slightly less than achieved by means of ultra-low-loss fibers and stronger input sidebands in \cite{Vanderhaegen2020OE}), each featuring a peak conversion followed by the return to the initial condition. The corresponding phase evolution is depicted in Fig.~\ref{exp_vs_num_Ps_new}(b) and shows nonlinear oscillations around the vanishing (input) phase, which remain bounded in the interval $\pm\pi/2$. This is further clear from the projection of the measured evolution in the phase-plane shown in Fig.~\ref{exp_vs_num_Ps_new}(c) for the sample value $\alpha_\mathrm{eff}=7~10^{-3}$ dB/km. As shown the trajectory remains bounded in the right semi-plane entailing repeated recurrences of the unshifted type. 

As the damping increases, a primary observation from Fig.~\ref{exp_vs_num_Ps_new}(a) is that the first peak conversion (or maximum compression in time) remains basically unchanged at $2.4$ km in the whole range of attenuations. This is in excellent agreement with the numerical simulations performed by integrating Eq.~(\ref{NLS}) and shown in Fig.~\ref{exp_vs_num_Ps_new}(f), and also fully supported by both perturbation approaches (dotted lines from Eq. \ref{alphac_IST} and see details in \cite{SM}). Indeed, the dissipation has no significant effect on the FPUT process in the first kilometers of the fiber. Conversely, an increasing attenuation has a strong and non-trivial impact on 
successive (second, third, and fourth) conversion peaks. Initially, at low enough $\alpha_\mathrm{eff}$, all such distances for peak conversion get longer when damping increases. Clearly, higher-order recurrences experience a larger rate of growth due to the accumulation of damping over longer distances. Remarkably, however, looking in particular at the third conversion peak,
its characteristic distance grows only up to a maximum, where it reaches nearly the fiber end. This condition is reached at the critical value of effective damping $\alpha_{eff,c2}=0.063$ dB/km corresponding to $P_R=205$ mW. In the whole range $\alpha_\mathrm{eff} \in [0,0.063]$ dB/km, denoted as region 3 in Fig.~\ref{exp_vs_num_Ps_new}(a), the first three recurrences are of the unshifted type (i.e. similar to the case in Fig.~\ref{exp_vs_num_Ps_new}(c)). 
However, if we keep on increasing  $\alpha_\mathrm{eff}$ beyond $\alpha_{eff,c2}=0.063$ dB/km, we notice an inverted trend where the distance of the third peak moderately decreases with increasing losses, as shown in Fig.~\ref{exp_vs_num_Ps_new}(a) for the experiment and in Fig.~\ref{exp_vs_num_Ps_new}(f) for the corresponding numerics.
Even more importantly, Fig.~\ref{exp_vs_num_Ps_new}(b) shows that, for $\alpha_\mathrm{eff}>\alpha_{eff,c2}$, the third peak conversion is characterized by a $\pi$ shift  of the relative phase.
This indicates that after two complete unshifted recurrences, the third one becomes of the shifted type. This is also consistently displayed by the phase-plane projection of the experimental data reported in Fig.~\ref{exp_vs_num_Ps_new}(d) (see also Fig.~\ref{exp_vs_num_Ps_new}(i) for the numerics), obtained for $\alpha_\mathrm{eff}=0.09$ dB/km. In this case, after the second return close to the origin which occurs around $Z=15$ km, the dynamics follows the double loop (shifted) trajectory, having crossed the separatrix. Importantly, this type of evolution characterizes the whole region 2 corresponding to $\alpha_{eff,c2} <\alpha_\mathrm{eff}<\alpha_{eff,c1}$.  Here, $\alpha_{eff,c1}=0.176$ dB/km (obtained at $P_R=55$ mW), is the characteristic damping where also the second conversion peak undergoes a symmetry breaking. Indeed, the second conversion peak is not critically affected by the previous transition occurring at $\alpha_{eff,c2}$, since its characteristic distance continues to increase smoothly for growing $\alpha_\mathrm{eff}$. However, across the damping $\alpha_\mathrm{eff}=\alpha_{eff,c1}$, this trend is clearly inverted (the distance starts to decrease for higher $\alpha_\mathrm{eff}$ above threshold $\alpha_{eff,c1}$), while a characteristic shift of $\pi$ appears in Fig.~\ref{exp_vs_num_Ps_new}(b,g) above threshold.
In this case, such transition is associated to the fact that, for $\alpha_\mathrm{eff}>\alpha_{eff,c1}$, the dynamics follows the shifted (double-loop) orbits with broken symmetry,
right after the first return close to the origin (i.e., the first recurrence). This is shown in Fig.~\ref{exp_vs_num_Ps_new}(e) (numerics in Fig.~\ref{exp_vs_num_Ps_new}(j)) for $\alpha_\mathrm{eff}=0.2$ dB/km. The net effect is that, in region 1 ($\alpha_\mathrm{eff}>\alpha_{eff,c1}$) only shifted recurrences can be followed, while repeated unshifted recurrences become inaccessible (see Fig.~S6 in the Supplementary).

Importantly, a good overall agreement is found between the experiment and the numerics in Fig.~\ref{exp_vs_num_Ps_new}, including the values of the critical losses
where $\alpha_{eff,c2}=0.063$ dB/km and $\alpha_{eff,c1}=0.176$ dB/km which compares with the numerical values $\alpha_{c2}=0.071$ dB/km and $\alpha_{c1}=0.142$ dB/km.

Finally we point out that the observation of critical transitions occurring at smaller values of damping and increasing number of recurrences as predicted from theory are prevented by the finite length of the fiber. In turn, operating with longer fiber and fine tuning of much smaller effective losses requires to improve the Raman amplification scheme, which will be addressed in the future.

{\it Conclusions}.  In summary, the accurate control of the effective fiber losses performed through Raman amplification techniques, has allowed us to demonstrate that the FPUT recurrence due to MI is strongly affected by damping. The main signature is the strong sensitivity of the recurrence periods around critical loss values where separatrix crossing is found to take place. On one hand, this is of crucial importance to design experiments aimed at investigating the transition to the thermalized state where FPUT recurrences break down.
On the other hand, our findings can be extended to other models, either integrable \cite{Ercolani1990_SGeq,Chen2020_Manakov,Liu2021_CLLeq} or strongly non-integrable \cite{Trillo1991_VNLS,Conforti2016}, which present separatrices and an underlying  geometric  structure of nonlinear MI. 
Finally, we also point out that similar phenomena of separatrix crossing (though, with inverted role of shifted-unshifted orbits) are envisaged when gain replaces losses \cite{Coppini2020}, which will be addressed in future experiments.

\section{Acknowledgments}
The present research was supported by IRCICA (USR 3380 CNRS), Agence Nationale de la Recherche (Programme Investissements d’Avenir, I-SITE VERIFICO); Ministry of Higher Education and Research; Hauts de France Council; European Regional Development Fund (Photonics for Society P4S, FELANI). S.T. acknowledges funding from Progetti di Ricerca di Interesse Nazionale (PRIN 2020X4T57A).

\bibliographystyle{apsrev4-1}
\bibliography{dampedFPUT_PRL}

\begin{thebibliography}{63}%
\makeatletter
\providecommand \@ifxundefined [1]{%
 \@ifx{#1\undefined}
}%
\providecommand \@ifnum [1]{%
 \ifnum #1\expandafter \@firstoftwo
 \else \expandafter \@secondoftwo
 \fi
}%
\providecommand \@ifx [1]{%
 \ifx #1\expandafter \@firstoftwo
 \else \expandafter \@secondoftwo
 \fi
}%
\providecommand \natexlab [1]{#1}%
\providecommand \enquote  [1]{``#1''}%
\providecommand \bibnamefont  [1]{#1}%
\providecommand \bibfnamefont [1]{#1}%
\providecommand \citenamefont [1]{#1}%
\providecommand \href@noop [0]{\@secondoftwo}%
\providecommand \href [0]{\begingroup \@sanitize@url \@href}%
\providecommand \@href[1]{\@@startlink{#1}\@@href}%
\providecommand \@@href[1]{\endgroup#1\@@endlink}%
\providecommand \@sanitize@url [0]{\catcode `\\12\catcode `\$12\catcode
  `\&12\catcode `\#12\catcode `\^12\catcode `\_12\catcode `\%12\relax}%
\providecommand \@@startlink[1]{}%
\providecommand \@@endlink[0]{}%
\providecommand \url  [0]{\begingroup\@sanitize@url \@url }%
\providecommand \@url [1]{\endgroup\@href {#1}{\urlprefix }}%
\providecommand \urlprefix  [0]{URL }%
\providecommand \Eprint [0]{\href }%
\providecommand \doibase [0]{http://dx.doi.org/}%
\providecommand \selectlanguage [0]{\@gobble}%
\providecommand \bibinfo  [0]{\@secondoftwo}%
\providecommand \bibfield  [0]{\@secondoftwo}%
\providecommand \translation [1]{[#1]}%
\providecommand \BibitemOpen [0]{}%
\providecommand \bibitemStop [0]{}%
\providecommand \bibitemNoStop [0]{.\EOS\space}%
\providecommand \EOS [0]{\spacefactor3000\relax}%
\providecommand \BibitemShut  [1]{\csname bibitem#1\endcsname}%
\let\auto@bib@innerbib\@empty
\bibitem [{\citenamefont {Fermi}\ \emph {et~al.}(1955)\citenamefont {Fermi},
  \citenamefont {Pasta}, \citenamefont {Ulam},\ and\ \citenamefont
  {Tsingou}}]{FPUToriginal}%
  \BibitemOpen
  \bibfield  {author} {\bibinfo {author} {\bibfnamefont {E.}~\bibnamefont
  {Fermi}}, \bibinfo {author} {\bibfnamefont {J.~R.}\ \bibnamefont {Pasta}},
  \bibinfo {author} {\bibfnamefont {S.}~\bibnamefont {Ulam}}, \ and\ \bibinfo
  {author} {\bibfnamefont {M.}~\bibnamefont {Tsingou}},\ }\href@noop {} {\emph
  {\bibinfo {title} {Studies of nonlinear problems, I,}}},\ \bibinfo {type}
  {Tech. Rep.}\ (\bibinfo  {institution} {Los Alamos Report No. LA-1940},\
  \bibinfo {year} {1955})\BibitemShut {NoStop}%
\bibitem [{\citenamefont {Ford}(1992)}]{Ford1992}%
  \BibitemOpen
  \bibfield  {author} {\bibinfo {author} {\bibfnamefont {J.}~\bibnamefont
  {Ford}},\ }\href {\doibase https://doi.org/10.1016/0370-1573(92)90116-H}
  {\bibfield  {journal} {\bibinfo  {journal} {Physics Reports}\ }\textbf
  {\bibinfo {volume} {213}},\ \bibinfo {pages} {271} (\bibinfo {year}
  {1992})}\BibitemShut {NoStop}%
\bibitem [{\citenamefont {Berman}\ and\ \citenamefont
  {Izrailev}(2005)}]{Berman2005}%
  \BibitemOpen
  \bibfield  {author} {\bibinfo {author} {\bibfnamefont {G.~P.}\ \bibnamefont
  {Berman}}\ and\ \bibinfo {author} {\bibfnamefont {F.~M.}\ \bibnamefont
  {Izrailev}},\ }\href {\doibase https://doi.org/10.1063/1.1855036} {\bibfield
  {journal} {\bibinfo  {journal} {Chaos}\ }\textbf {\bibinfo {volume} {15}},\
  \bibinfo {pages} {015104} (\bibinfo {year} {2005})}\BibitemShut {NoStop}%
\bibitem [{\citenamefont {Trillo}\ \emph {et~al.}(2016)\citenamefont {Trillo},
  \citenamefont {Deng}, \citenamefont {Biondini}, \citenamefont {Klein},
  \citenamefont {Clauss}, \citenamefont {Chabchoub},\ and\ \citenamefont
  {Onorato}}]{Trillo2016}%
  \BibitemOpen
  \bibfield  {author} {\bibinfo {author} {\bibfnamefont {S.}~\bibnamefont
  {Trillo}}, \bibinfo {author} {\bibfnamefont {G.}~\bibnamefont {Deng}},
  \bibinfo {author} {\bibfnamefont {G.}~\bibnamefont {Biondini}}, \bibinfo
  {author} {\bibfnamefont {M.}~\bibnamefont {Klein}}, \bibinfo {author}
  {\bibfnamefont {G.~F.}\ \bibnamefont {Clauss}}, \bibinfo {author}
  {\bibfnamefont {A.}~\bibnamefont {Chabchoub}}, \ and\ \bibinfo {author}
  {\bibfnamefont {M.}~\bibnamefont {Onorato}},\ }\href {\doibase
  10.1103/PhysRevLett.117.144102} {\bibfield  {journal} {\bibinfo  {journal}
  {Phys. Rev. Lett.}\ }\textbf {\bibinfo {volume} {117}},\ \bibinfo {pages}
  {144102} (\bibinfo {year} {2016})}\BibitemShut {NoStop}%
\bibitem [{\citenamefont {Guasoni}\ \emph {et~al.}(2017)\citenamefont
  {Guasoni}, \citenamefont {Garnier}, \citenamefont {Rumpf}, \citenamefont
  {Sugny}, \citenamefont {Fatome}, \citenamefont {Amrani}, \citenamefont
  {Millot},\ and\ \citenamefont {Picozzi}}]{Guasoni2017}%
  \BibitemOpen
  \bibfield  {author} {\bibinfo {author} {\bibfnamefont {M.}~\bibnamefont
  {Guasoni}}, \bibinfo {author} {\bibfnamefont {J.}~\bibnamefont {Garnier}},
  \bibinfo {author} {\bibfnamefont {B.}~\bibnamefont {Rumpf}}, \bibinfo
  {author} {\bibfnamefont {D.}~\bibnamefont {Sugny}}, \bibinfo {author}
  {\bibfnamefont {J.}~\bibnamefont {Fatome}}, \bibinfo {author} {\bibfnamefont
  {F.}~\bibnamefont {Amrani}}, \bibinfo {author} {\bibfnamefont
  {G.}~\bibnamefont {Millot}}, \ and\ \bibinfo {author} {\bibfnamefont
  {A.}~\bibnamefont {Picozzi}},\ }\href {\doibase 10.1103/PhysRevX.7.011025}
  {\bibfield  {journal} {\bibinfo  {journal} {Phys. Rev. X}\ }\textbf {\bibinfo
  {volume} {7}},\ \bibinfo {pages} {011025} (\bibinfo {year}
  {2017})}\BibitemShut {NoStop}%
\bibitem [{\citenamefont {Lvov}\ and\ \citenamefont
  {Onorato}(2018)}]{Lvov2018}%
  \BibitemOpen
  \bibfield  {author} {\bibinfo {author} {\bibfnamefont {Y.~V.}\ \bibnamefont
  {Lvov}}\ and\ \bibinfo {author} {\bibfnamefont {M.}~\bibnamefont {Onorato}},\
  }\href {\doibase 10.1103/PhysRevLett.120.144301} {\bibfield  {journal}
  {\bibinfo  {journal} {Phys. Rev. Lett.}\ }\textbf {\bibinfo {volume} {120}},\
  \bibinfo {pages} {144301} (\bibinfo {year} {2018})}\BibitemShut {NoStop}%
\bibitem [{\citenamefont {Pace}\ \emph {et~al.}(2019)\citenamefont {Pace},
  \citenamefont {Reiss},\ and\ \citenamefont {Campbell}}]{Pace2019b}%
  \BibitemOpen
  \bibfield  {author} {\bibinfo {author} {\bibfnamefont {S.~D.}\ \bibnamefont
  {Pace}}, \bibinfo {author} {\bibfnamefont {K.~A.}\ \bibnamefont {Reiss}}, \
  and\ \bibinfo {author} {\bibfnamefont {D.~K.}\ \bibnamefont {Campbell}},\
  }\href {\doibase 10.1063/1.5122972} {\bibfield  {journal} {\bibinfo
  {journal} {Chaos}\ }\textbf {\bibinfo {volume} {29}},\ \bibinfo {pages}
  {113107} (\bibinfo {year} {2019})}\BibitemShut {NoStop}%
\bibitem [{\citenamefont {Zaleski}\ \emph {et~al.}(2020)\citenamefont
  {Zaleski}, \citenamefont {Onorato},\ and\ \citenamefont
  {Lvov}}]{Zaleski2020}%
  \BibitemOpen
  \bibfield  {author} {\bibinfo {author} {\bibfnamefont {J.}~\bibnamefont
  {Zaleski}}, \bibinfo {author} {\bibfnamefont {M.}~\bibnamefont {Onorato}}, \
  and\ \bibinfo {author} {\bibfnamefont {Y.~V.}\ \bibnamefont {Lvov}},\ }\href
  {\doibase 10.1103/PhysRevX.10.021043} {\bibfield  {journal} {\bibinfo
  {journal} {Phys. Rev. X}\ }\textbf {\bibinfo {volume} {10}},\ \bibinfo
  {pages} {021043} (\bibinfo {year} {2020})}\BibitemShut {NoStop}%
\bibitem [{\citenamefont {Dematteis}\ \emph {et~al.}(2020)\citenamefont
  {Dematteis}, \citenamefont {Rondoni}, \citenamefont {Proment}, \citenamefont
  {De~Vita},\ and\ \citenamefont {Onorato}}]{Dematteis2020}%
  \BibitemOpen
  \bibfield  {author} {\bibinfo {author} {\bibfnamefont {G.}~\bibnamefont
  {Dematteis}}, \bibinfo {author} {\bibfnamefont {L.}~\bibnamefont {Rondoni}},
  \bibinfo {author} {\bibfnamefont {D.}~\bibnamefont {Proment}}, \bibinfo
  {author} {\bibfnamefont {F.}~\bibnamefont {De~Vita}}, \ and\ \bibinfo
  {author} {\bibfnamefont {M.}~\bibnamefont {Onorato}},\ }\href {\doibase
  10.1103/PhysRevLett.125.024101} {\bibfield  {journal} {\bibinfo  {journal}
  {Phys. Rev. Lett.}\ }\textbf {\bibinfo {volume} {125}},\ \bibinfo {pages}
  {024101} (\bibinfo {year} {2020})}\BibitemShut {NoStop}%
\bibitem [{\citenamefont {Wabnitz}\ and\ \citenamefont
  {Wetzel}(2014)}]{Wabnitz2014}%
  \BibitemOpen
  \bibfield  {author} {\bibinfo {author} {\bibfnamefont {S.}~\bibnamefont
  {Wabnitz}}\ and\ \bibinfo {author} {\bibfnamefont {B.}~\bibnamefont
  {Wetzel}},\ }\href {\doibase https://doi.org/10.1016/j.physleta.2014.07.018}
  {\bibfield  {journal} {\bibinfo  {journal} {Phys. Lett. A}\ }\textbf
  {\bibinfo {volume} {378}},\ \bibinfo {pages} {2750} (\bibinfo {year}
  {2014})}\BibitemShut {NoStop}%
\bibitem [{\citenamefont {Zakharov}\ and\ \citenamefont
  {Ostrovsky}(2009)}]{ZO2009}%
  \BibitemOpen
  \bibfield  {author} {\bibinfo {author} {\bibfnamefont {V.}~\bibnamefont
  {Zakharov}}\ and\ \bibinfo {author} {\bibfnamefont {L.}~\bibnamefont
  {Ostrovsky}},\ }\href {\doibase
  http://dx.doi.org/10.1016/j.physd.2008.12.002} {\bibfield  {journal}
  {\bibinfo  {journal} {Physica D}\ }\textbf {\bibinfo {volume} {238}},\
  \bibinfo {pages} {540} (\bibinfo {year} {2009})}\BibitemShut {NoStop}%
\bibitem [{\citenamefont {Benjamin}\ and\ \citenamefont {Feir}(1967)}]{BF1967}%
  \BibitemOpen
  \bibfield  {author} {\bibinfo {author} {\bibfnamefont {T.~B.}\ \bibnamefont
  {Benjamin}}\ and\ \bibinfo {author} {\bibfnamefont {J.~E.}\ \bibnamefont
  {Feir}},\ }\href {\doibase 10.1017/S002211206700045X} {\bibfield  {journal}
  {\bibinfo  {journal} {Journal of Fluid Mechanics}\ }\textbf {\bibinfo
  {volume} {27}},\ \bibinfo {pages} {417–430} (\bibinfo {year}
  {1967})}\BibitemShut {NoStop}%
\bibitem [{\citenamefont {Taniuti}\ and\ \citenamefont
  {Washimi}(1968)}]{Taniuti1968}%
  \BibitemOpen
  \bibfield  {author} {\bibinfo {author} {\bibfnamefont {T.}~\bibnamefont
  {Taniuti}}\ and\ \bibinfo {author} {\bibfnamefont {H.}~\bibnamefont
  {Washimi}},\ }\href {\doibase 10.1103/PhysRevLett.21.209} {\bibfield
  {journal} {\bibinfo  {journal} {Phys. Rev. Lett.}\ }\textbf {\bibinfo
  {volume} {21}},\ \bibinfo {pages} {209} (\bibinfo {year} {1968})}\BibitemShut
  {NoStop}%
\bibitem [{\citenamefont {Tai}\ \emph {et~al.}(1986)\citenamefont {Tai},
  \citenamefont {Hasegawa},\ and\ \citenamefont {Tomita}}]{Tai1986}%
  \BibitemOpen
  \bibfield  {author} {\bibinfo {author} {\bibfnamefont {K.}~\bibnamefont
  {Tai}}, \bibinfo {author} {\bibfnamefont {A.}~\bibnamefont {Hasegawa}}, \
  and\ \bibinfo {author} {\bibfnamefont {A.}~\bibnamefont {Tomita}},\ }\href
  {\doibase 10.1103/PhysRevLett.56.135} {\bibfield  {journal} {\bibinfo
  {journal} {Phys. Rev. Lett.}\ }\textbf {\bibinfo {volume} {56}},\ \bibinfo
  {pages} {135} (\bibinfo {year} {1986})}\BibitemShut {NoStop}%
\bibitem [{\citenamefont {Everitt}\ \emph {et~al.}(2017)\citenamefont
  {Everitt}, \citenamefont {Sooriyabandara}, \citenamefont {Guasoni},
  \citenamefont {Wigley}, \citenamefont {Wei}, \citenamefont {McDonald},
  \citenamefont {Hardman}, \citenamefont {Manju}, \citenamefont {Close},
  \citenamefont {Kuhn}, \citenamefont {Szigeti}, \citenamefont {Kivshar},\ and\
  \citenamefont {Robins}}]{Everitt2017}%
  \BibitemOpen
  \bibfield  {author} {\bibinfo {author} {\bibfnamefont {P.~J.}\ \bibnamefont
  {Everitt}}, \bibinfo {author} {\bibfnamefont {M.~A.}\ \bibnamefont
  {Sooriyabandara}}, \bibinfo {author} {\bibfnamefont {M.}~\bibnamefont
  {Guasoni}}, \bibinfo {author} {\bibfnamefont {P.~B.}\ \bibnamefont {Wigley}},
  \bibinfo {author} {\bibfnamefont {C.~H.}\ \bibnamefont {Wei}}, \bibinfo
  {author} {\bibfnamefont {G.~D.}\ \bibnamefont {McDonald}}, \bibinfo {author}
  {\bibfnamefont {K.~S.}\ \bibnamefont {Hardman}}, \bibinfo {author}
  {\bibfnamefont {P.}~\bibnamefont {Manju}}, \bibinfo {author} {\bibfnamefont
  {J.~D.}\ \bibnamefont {Close}}, \bibinfo {author} {\bibfnamefont {C.~C.~N.}\
  \bibnamefont {Kuhn}}, \bibinfo {author} {\bibfnamefont {S.~S.}\ \bibnamefont
  {Szigeti}}, \bibinfo {author} {\bibfnamefont {Y.~S.}\ \bibnamefont
  {Kivshar}}, \ and\ \bibinfo {author} {\bibfnamefont {N.~P.}\ \bibnamefont
  {Robins}},\ }\href {\doibase 10.1103/PhysRevA.96.041601} {\bibfield
  {journal} {\bibinfo  {journal} {Phys. Rev. A}\ }\textbf {\bibinfo {volume}
  {96}},\ \bibinfo {pages} {041601} (\bibinfo {year} {2017})}\BibitemShut
  {NoStop}%
\bibitem [{\citenamefont {Leykam}\ \emph {et~al.}(2021)\citenamefont {Leykam},
  \citenamefont {Smolina}, \citenamefont {Maluckov}, \citenamefont {Flach},\
  and\ \citenamefont {Smirnova}}]{Leykam2021}%
  \BibitemOpen
  \bibfield  {author} {\bibinfo {author} {\bibfnamefont {D.}~\bibnamefont
  {Leykam}}, \bibinfo {author} {\bibfnamefont {E.}~\bibnamefont {Smolina}},
  \bibinfo {author} {\bibfnamefont {A.}~\bibnamefont {Maluckov}}, \bibinfo
  {author} {\bibfnamefont {S.}~\bibnamefont {Flach}}, \ and\ \bibinfo {author}
  {\bibfnamefont {D.~A.}\ \bibnamefont {Smirnova}},\ }\href {\doibase
  10.1103/PhysRevLett.126.073901} {\bibfield  {journal} {\bibinfo  {journal}
  {Phys. Rev. Lett.}\ }\textbf {\bibinfo {volume} {126}},\ \bibinfo {pages}
  {073901} (\bibinfo {year} {2021})}\BibitemShut {NoStop}%
\bibitem [{\citenamefont {Erkintalo}\ \emph {et~al.}(2014)\citenamefont
  {Erkintalo}, \citenamefont {Dias}, \citenamefont {Dudley},\ and\
  \citenamefont {Genty}}]{Dudley2014}%
  \BibitemOpen
  \bibfield  {author} {\bibinfo {author} {\bibfnamefont {M.}~\bibnamefont
  {Erkintalo}}, \bibinfo {author} {\bibfnamefont {F.}~\bibnamefont {Dias}},
  \bibinfo {author} {\bibfnamefont {J.~M.}\ \bibnamefont {Dudley}}, \ and\
  \bibinfo {author} {\bibfnamefont {G.}~\bibnamefont {Genty}},\ }\href@noop {}
  {\bibfield  {journal} {\bibinfo  {journal} {Nat. Photonics}\ }\textbf
  {\bibinfo {volume} {8}},\ \bibinfo {pages} {755} (\bibinfo {year}
  {2014})}\BibitemShut {NoStop}%
\bibitem [{\citenamefont {Onorato}\ \emph {et~al.}(2013)\citenamefont
  {Onorato}, \citenamefont {Residori}, \citenamefont {Bortolozzo},
  \citenamefont {Montina},\ and\ \citenamefont {Arecchi}}]{Onorato2013}%
  \BibitemOpen
  \bibfield  {author} {\bibinfo {author} {\bibfnamefont {M.}~\bibnamefont
  {Onorato}}, \bibinfo {author} {\bibfnamefont {S.}~\bibnamefont {Residori}},
  \bibinfo {author} {\bibfnamefont {U.}~\bibnamefont {Bortolozzo}}, \bibinfo
  {author} {\bibfnamefont {A.}~\bibnamefont {Montina}}, \ and\ \bibinfo
  {author} {\bibfnamefont {F.}~\bibnamefont {Arecchi}},\ }\href {\doibase
  https://doi.org/10.1016/j.physrep.2013.03.001} {\bibfield  {journal}
  {\bibinfo  {journal} {Phys. Rep.}\ }\textbf {\bibinfo {volume} {528}},\
  \bibinfo {pages} {47} (\bibinfo {year} {2013})}\BibitemShut {NoStop}%
\bibitem [{\citenamefont {Zakharov}\ and\ \citenamefont
  {Gelash}(2013)}]{ZG2013}%
  \BibitemOpen
  \bibfield  {author} {\bibinfo {author} {\bibfnamefont {V.~E.}\ \bibnamefont
  {Zakharov}}\ and\ \bibinfo {author} {\bibfnamefont {A.~A.}\ \bibnamefont
  {Gelash}},\ }\href {\doibase 10.1103/PhysRevLett.111.054101} {\bibfield
  {journal} {\bibinfo  {journal} {Phys. Rev. Lett.}\ }\textbf {\bibinfo
  {volume} {111}},\ \bibinfo {pages} {054101} (\bibinfo {year}
  {2013})}\BibitemShut {NoStop}%
\bibitem [{\citenamefont {Biondini}\ and\ \citenamefont
  {Mantzavinos}(2016)}]{Biondini2016}%
  \BibitemOpen
  \bibfield  {author} {\bibinfo {author} {\bibfnamefont {G.}~\bibnamefont
  {Biondini}}\ and\ \bibinfo {author} {\bibfnamefont {D.}~\bibnamefont
  {Mantzavinos}},\ }\href {\doibase 10.1103/PhysRevLett.116.043902} {\bibfield
  {journal} {\bibinfo  {journal} {Phys. Rev. Lett.}\ }\textbf {\bibinfo
  {volume} {116}},\ \bibinfo {pages} {043902} (\bibinfo {year}
  {2016})}\BibitemShut {NoStop}%
\bibitem [{\citenamefont {Soto-Crespo}\ \emph {et~al.}(2016)\citenamefont
  {Soto-Crespo}, \citenamefont {Devine},\ and\ \citenamefont
  {Akhmediev}}]{SotoCrespo2016}%
  \BibitemOpen
  \bibfield  {author} {\bibinfo {author} {\bibfnamefont {J.~M.}\ \bibnamefont
  {Soto-Crespo}}, \bibinfo {author} {\bibfnamefont {N.}~\bibnamefont {Devine}},
  \ and\ \bibinfo {author} {\bibfnamefont {N.}~\bibnamefont {Akhmediev}},\
  }\href {\doibase 10.1103/PhysRevLett.116.103901} {\bibfield  {journal}
  {\bibinfo  {journal} {Phys. Rev. Lett.}\ }\textbf {\bibinfo {volume} {116}},\
  \bibinfo {pages} {103901} (\bibinfo {year} {2016})}\BibitemShut {NoStop}%
\bibitem [{\citenamefont {Armaroli}\ \emph {et~al.}(2017)\citenamefont
  {Armaroli}, \citenamefont {Brunetti},\ and\ \citenamefont
  {Kasparian}}]{Armaroli_2017}%
  \BibitemOpen
  \bibfield  {author} {\bibinfo {author} {\bibfnamefont {A.}~\bibnamefont
  {Armaroli}}, \bibinfo {author} {\bibfnamefont {M.}~\bibnamefont {Brunetti}},
  \ and\ \bibinfo {author} {\bibfnamefont {J.}~\bibnamefont {Kasparian}},\
  }\href {\doibase 10.1103/PhysRevE.96.012222} {\bibfield  {journal} {\bibinfo
  {journal} {Physical Review E}\ }\textbf {\bibinfo {volume} {96}},\ \bibinfo
  {pages} {012222} (\bibinfo {year} {2017})}\BibitemShut {NoStop}%
\bibitem [{\citenamefont {Armaroli}\ \emph {et~al.}(2018)\citenamefont
  {Armaroli}, \citenamefont {Eeltink}, \citenamefont {Brunetti},\ and\
  \citenamefont {Kasparian}}]{Armaroli_2018}%
  \BibitemOpen
  \bibfield  {author} {\bibinfo {author} {\bibfnamefont {A.}~\bibnamefont
  {Armaroli}}, \bibinfo {author} {\bibfnamefont {D.}~\bibnamefont {Eeltink}},
  \bibinfo {author} {\bibfnamefont {M.}~\bibnamefont {Brunetti}}, \ and\
  \bibinfo {author} {\bibfnamefont {J.}~\bibnamefont {Kasparian}},\ }\href
  {\doibase 10.1063/1.5006139} {\bibfield  {journal} {\bibinfo  {journal}
  {Physics of Fluids}\ }\textbf {\bibinfo {volume} {30}},\ \bibinfo {pages}
  {017102} (\bibinfo {year} {2018})}\BibitemShut {NoStop}%
\bibitem [{\citenamefont {Grinevich}\ and\ \citenamefont
  {Santini}(2018{\natexlab{a}})}]{GS2018a}%
  \BibitemOpen
  \bibfield  {author} {\bibinfo {author} {\bibfnamefont {P.}~\bibnamefont
  {Grinevich}}\ and\ \bibinfo {author} {\bibfnamefont {P.}~\bibnamefont
  {Santini}},\ }\href {\doibase https://doi.org/10.1016/j.physleta.2018.02.014}
  {\bibfield  {journal} {\bibinfo  {journal} {Phys. Lett. A}\ }\textbf
  {\bibinfo {volume} {382}},\ \bibinfo {pages} {973} (\bibinfo {year}
  {2018}{\natexlab{a}})}\BibitemShut {NoStop}%
\bibitem [{\citenamefont {Grinevich}\ and\ \citenamefont
  {Santini}(2018{\natexlab{b}})}]{GS2018b}%
  \BibitemOpen
  \bibfield  {author} {\bibinfo {author} {\bibfnamefont {P.~G.}\ \bibnamefont
  {Grinevich}}\ and\ \bibinfo {author} {\bibfnamefont {P.~M.}\ \bibnamefont
  {Santini}},\ }\href {\doibase 10.1088/1361-6544/aaddcf} {\bibfield  {journal}
  {\bibinfo  {journal} {Nonlinearity}\ }\textbf {\bibinfo {volume} {31}},\
  \bibinfo {pages} {5258} (\bibinfo {year} {2018}{\natexlab{b}})}\BibitemShut
  {NoStop}%
\bibitem [{\citenamefont {Biondini}\ \emph {et~al.}(2018)\citenamefont
  {Biondini}, \citenamefont {Li}, \citenamefont {Mantzavinos},\ and\
  \citenamefont {Trillo}}]{Biondini2018}%
  \BibitemOpen
  \bibfield  {author} {\bibinfo {author} {\bibfnamefont {G.}~\bibnamefont
  {Biondini}}, \bibinfo {author} {\bibfnamefont {S.}~\bibnamefont {Li}},
  \bibinfo {author} {\bibfnamefont {D.}~\bibnamefont {Mantzavinos}}, \ and\
  \bibinfo {author} {\bibfnamefont {S.}~\bibnamefont {Trillo}},\ }\href
  {\doibase 10.1137/17M1112765} {\bibfield  {journal} {\bibinfo  {journal}
  {SIAM Rev.}\ }\textbf {\bibinfo {volume} {60}},\ \bibinfo {pages} {888}
  (\bibinfo {year} {2018})}\BibitemShut {NoStop}%
\bibitem [{\citenamefont {Conforti}\ \emph {et~al.}(2018)\citenamefont
  {Conforti}, \citenamefont {Li}, \citenamefont {Biondini},\ and\ \citenamefont
  {Trillo}}]{Conforti2018}%
  \BibitemOpen
  \bibfield  {author} {\bibinfo {author} {\bibfnamefont {M.}~\bibnamefont
  {Conforti}}, \bibinfo {author} {\bibfnamefont {S.}~\bibnamefont {Li}},
  \bibinfo {author} {\bibfnamefont {G.}~\bibnamefont {Biondini}}, \ and\
  \bibinfo {author} {\bibfnamefont {S.}~\bibnamefont {Trillo}},\ }\href
  {\doibase 10.1364/OL.43.005291} {\bibfield  {journal} {\bibinfo  {journal}
  {Opt. Lett.}\ }\textbf {\bibinfo {volume} {43}},\ \bibinfo {pages} {5291}
  (\bibinfo {year} {2018})}\BibitemShut {NoStop}%
\bibitem [{\citenamefont {Trillo}\ and\ \citenamefont
  {Conforti}(2019)}]{Trillo2019}%
  \BibitemOpen
  \bibfield  {author} {\bibinfo {author} {\bibfnamefont {S.}~\bibnamefont
  {Trillo}}\ and\ \bibinfo {author} {\bibfnamefont {M.}~\bibnamefont
  {Conforti}},\ }\href {\doibase 10.1364/OL.44.004275} {\bibfield  {journal}
  {\bibinfo  {journal} {Opt. Lett.}\ }\textbf {\bibinfo {volume} {44}},\
  \bibinfo {pages} {4275} (\bibinfo {year} {2019})}\BibitemShut {NoStop}%
\bibitem [{\citenamefont {Conforti}\ \emph {et~al.}(2020)\citenamefont
  {Conforti}, \citenamefont {Mussot}, \citenamefont {Kudlinski}, \citenamefont
  {Trillo},\ and\ \citenamefont {Akhmediev}}]{Conforti2020}%
  \BibitemOpen
  \bibfield  {author} {\bibinfo {author} {\bibfnamefont {M.}~\bibnamefont
  {Conforti}}, \bibinfo {author} {\bibfnamefont {A.}~\bibnamefont {Mussot}},
  \bibinfo {author} {\bibfnamefont {A.}~\bibnamefont {Kudlinski}}, \bibinfo
  {author} {\bibfnamefont {S.}~\bibnamefont {Trillo}}, \ and\ \bibinfo {author}
  {\bibfnamefont {N.}~\bibnamefont {Akhmediev}},\ }\href {\doibase
  10.1103/PhysRevA.101.023843} {\bibfield  {journal} {\bibinfo  {journal}
  {Physical Review A}\ }\textbf {\bibinfo {volume} {101}},\ \bibinfo {pages}
  {023843} (\bibinfo {year} {2020})}\BibitemShut {NoStop}%
\bibitem [{\citenamefont {Van~Simaeys}\ \emph {et~al.}(2001)\citenamefont
  {Van~Simaeys}, \citenamefont {Emplit},\ and\ \citenamefont
  {Haelterman}}]{VanSimaeys2001}%
  \BibitemOpen
  \bibfield  {author} {\bibinfo {author} {\bibfnamefont {G.}~\bibnamefont
  {Van~Simaeys}}, \bibinfo {author} {\bibfnamefont {P.}~\bibnamefont {Emplit}},
  \ and\ \bibinfo {author} {\bibfnamefont {M.}~\bibnamefont {Haelterman}},\
  }\href {\doibase 10.1103/PhysRevLett.87.033902} {\bibfield  {journal}
  {\bibinfo  {journal} {Phys. Rev. Lett.}\ }\textbf {\bibinfo {volume} {87}},\
  \bibinfo {pages} {033902} (\bibinfo {year} {2001})}\BibitemShut {NoStop}%
\bibitem [{\citenamefont {Erkintalo}\ \emph {et~al.}(2011)\citenamefont
  {Erkintalo}, \citenamefont {Hammani}, \citenamefont {Kibler}, \citenamefont
  {Finot}, \citenamefont {Akhmediev}, \citenamefont {Dudley},\ and\
  \citenamefont {Genty}}]{Erkintalo2011}%
  \BibitemOpen
  \bibfield  {author} {\bibinfo {author} {\bibfnamefont {M.}~\bibnamefont
  {Erkintalo}}, \bibinfo {author} {\bibfnamefont {K.}~\bibnamefont {Hammani}},
  \bibinfo {author} {\bibfnamefont {B.}~\bibnamefont {Kibler}}, \bibinfo
  {author} {\bibfnamefont {C.}~\bibnamefont {Finot}}, \bibinfo {author}
  {\bibfnamefont {N.}~\bibnamefont {Akhmediev}}, \bibinfo {author}
  {\bibfnamefont {J.~M.}\ \bibnamefont {Dudley}}, \ and\ \bibinfo {author}
  {\bibfnamefont {G.}~\bibnamefont {Genty}},\ }\href {\doibase
  10.1103/PhysRevLett.107.253901} {\bibfield  {journal} {\bibinfo  {journal}
  {Phys. Rev. Lett.}\ }\textbf {\bibinfo {volume} {107}},\ \bibinfo {pages}
  {253901} (\bibinfo {year} {2011})}\BibitemShut {NoStop}%
\bibitem [{\citenamefont {Mussot}\ \emph {et~al.}(2014)\citenamefont {Mussot},
  \citenamefont {Kudlinski}, \citenamefont {Droques}, \citenamefont
  {Szriftgiser},\ and\ \citenamefont {Akhmediev}}]{Mussot2014}%
  \BibitemOpen
  \bibfield  {author} {\bibinfo {author} {\bibfnamefont {A.}~\bibnamefont
  {Mussot}}, \bibinfo {author} {\bibfnamefont {A.}~\bibnamefont {Kudlinski}},
  \bibinfo {author} {\bibfnamefont {M.}~\bibnamefont {Droques}}, \bibinfo
  {author} {\bibfnamefont {P.}~\bibnamefont {Szriftgiser}}, \ and\ \bibinfo
  {author} {\bibfnamefont {N.}~\bibnamefont {Akhmediev}},\ }\href {\doibase
  10.1103/PhysRevX.4.011054} {\bibfield  {journal} {\bibinfo  {journal} {Phys.
  Rev. X}\ }\textbf {\bibinfo {volume} {4}},\ \bibinfo {pages} {011054}
  (\bibinfo {year} {2014})}\BibitemShut {NoStop}%
\bibitem [{\citenamefont {Kibler}\ \emph {et~al.}(2015)\citenamefont {Kibler},
  \citenamefont {Chabchoub}, \citenamefont {Gelash}, \citenamefont
  {Akhmediev},\ and\ \citenamefont {Zakharov}}]{Kibler2015}%
  \BibitemOpen
  \bibfield  {author} {\bibinfo {author} {\bibfnamefont {B.}~\bibnamefont
  {Kibler}}, \bibinfo {author} {\bibfnamefont {A.}~\bibnamefont {Chabchoub}},
  \bibinfo {author} {\bibfnamefont {A.}~\bibnamefont {Gelash}}, \bibinfo
  {author} {\bibfnamefont {N.}~\bibnamefont {Akhmediev}}, \ and\ \bibinfo
  {author} {\bibfnamefont {V.~E.}\ \bibnamefont {Zakharov}},\ }\href {\doibase
  10.1103/PhysRevX.5.041026} {\bibfield  {journal} {\bibinfo  {journal} {Phys.
  Rev. X}\ }\textbf {\bibinfo {volume} {5}},\ \bibinfo {pages} {041026}
  (\bibinfo {year} {2015})}\BibitemShut {NoStop}%
\bibitem [{\citenamefont {Mussot}\ \emph {et~al.}(2018)\citenamefont {Mussot},
  \citenamefont {Naveau}, \citenamefont {Conforti}, \citenamefont {Kudlinski},
  \citenamefont {Copie}, \citenamefont {Szriftgiser},\ and\ \citenamefont
  {Trillo}}]{Mussot2018}%
  \BibitemOpen
  \bibfield  {author} {\bibinfo {author} {\bibfnamefont {A.}~\bibnamefont
  {Mussot}}, \bibinfo {author} {\bibfnamefont {C.}~\bibnamefont {Naveau}},
  \bibinfo {author} {\bibfnamefont {M.}~\bibnamefont {Conforti}}, \bibinfo
  {author} {\bibfnamefont {A.}~\bibnamefont {Kudlinski}}, \bibinfo {author}
  {\bibfnamefont {F.}~\bibnamefont {Copie}}, \bibinfo {author} {\bibfnamefont
  {P.}~\bibnamefont {Szriftgiser}}, \ and\ \bibinfo {author} {\bibfnamefont
  {S.}~\bibnamefont {Trillo}},\ }\href {\doibase 10.1038/s41566-018-0136-1}
  {\bibfield  {journal} {\bibinfo  {journal} {Nat. Photonics}\ }\textbf
  {\bibinfo {volume} {12}},\ \bibinfo {pages} {303} (\bibinfo {year}
  {2018})}\BibitemShut {NoStop}%
\bibitem [{\citenamefont {Pierangeli}\ \emph {et~al.}(2018)\citenamefont
  {Pierangeli}, \citenamefont {Flammini}, \citenamefont {Zhang}, \citenamefont
  {Marcucci}, \citenamefont {Agranat}, \citenamefont {Grinevich}, \citenamefont
  {Santini}, \citenamefont {Conti},\ and\ \citenamefont
  {DelRe}}]{Pierangeli2018}%
  \BibitemOpen
  \bibfield  {author} {\bibinfo {author} {\bibfnamefont {D.}~\bibnamefont
  {Pierangeli}}, \bibinfo {author} {\bibfnamefont {M.}~\bibnamefont
  {Flammini}}, \bibinfo {author} {\bibfnamefont {L.}~\bibnamefont {Zhang}},
  \bibinfo {author} {\bibfnamefont {G.}~\bibnamefont {Marcucci}}, \bibinfo
  {author} {\bibfnamefont {A.~J.}\ \bibnamefont {Agranat}}, \bibinfo {author}
  {\bibfnamefont {P.~G.}\ \bibnamefont {Grinevich}}, \bibinfo {author}
  {\bibfnamefont {P.~M.}\ \bibnamefont {Santini}}, \bibinfo {author}
  {\bibfnamefont {C.}~\bibnamefont {Conti}}, \ and\ \bibinfo {author}
  {\bibfnamefont {E.}~\bibnamefont {DelRe}},\ }\href {\doibase
  10.1103/PhysRevX.8.041017} {\bibfield  {journal} {\bibinfo  {journal} {Phys.
  Rev. X}\ }\textbf {\bibinfo {volume} {8}},\ \bibinfo {pages} {041017}
  (\bibinfo {year} {2018})}\BibitemShut {NoStop}%
\bibitem [{\citenamefont {Kraych}\ \emph
  {et~al.}(2019{\natexlab{a}})\citenamefont {Kraych}, \citenamefont {Suret},
  \citenamefont {El},\ and\ \citenamefont {Randoux}}]{Kraych2019a}%
  \BibitemOpen
  \bibfield  {author} {\bibinfo {author} {\bibfnamefont {A.~E.}\ \bibnamefont
  {Kraych}}, \bibinfo {author} {\bibfnamefont {P.}~\bibnamefont {Suret}},
  \bibinfo {author} {\bibfnamefont {G.}~\bibnamefont {El}}, \ and\ \bibinfo
  {author} {\bibfnamefont {S.}~\bibnamefont {Randoux}},\ }\href {\doibase
  10.1103/PhysRevLett.122.054101} {\bibfield  {journal} {\bibinfo  {journal}
  {Phys. Rev. Lett.}\ }\textbf {\bibinfo {volume} {122}},\ \bibinfo {pages}
  {054101} (\bibinfo {year} {2019}{\natexlab{a}})}\BibitemShut {NoStop}%
\bibitem [{\citenamefont {Kraych}\ \emph
  {et~al.}(2019{\natexlab{b}})\citenamefont {Kraych}, \citenamefont
  {Agafontsev}, \citenamefont {Randoux},\ and\ \citenamefont
  {Suret}}]{Kraych2019b}%
  \BibitemOpen
  \bibfield  {author} {\bibinfo {author} {\bibfnamefont {A.~E.}\ \bibnamefont
  {Kraych}}, \bibinfo {author} {\bibfnamefont {D.}~\bibnamefont {Agafontsev}},
  \bibinfo {author} {\bibfnamefont {S.}~\bibnamefont {Randoux}}, \ and\
  \bibinfo {author} {\bibfnamefont {P.}~\bibnamefont {Suret}},\ }\href
  {\doibase 10.1103/PhysRevLett.123.093902} {\bibfield  {journal} {\bibinfo
  {journal} {Phys. Rev. Lett.}\ }\textbf {\bibinfo {volume} {123}},\ \bibinfo
  {pages} {093902} (\bibinfo {year} {2019}{\natexlab{b}})}\BibitemShut
  {NoStop}%
\bibitem [{\citenamefont {Goossens}\ \emph {et~al.}(2019)\citenamefont
  {Goossens}, \citenamefont {Hafermann},\ and\ \citenamefont
  {Jaou\"en}}]{Goossens2019}%
  \BibitemOpen
  \bibfield  {author} {\bibinfo {author} {\bibfnamefont {J.-W.}\ \bibnamefont
  {Goossens}}, \bibinfo {author} {\bibfnamefont {H.}~\bibnamefont {Hafermann}},
  \ and\ \bibinfo {author} {\bibfnamefont {Y.}~\bibnamefont {Jaou\"en}},\
  }\href {\doibase 10.1038/s41598-019-54825-4} {\bibfield  {journal} {\bibinfo
  {journal} {Sci. Rep.}\ }\textbf {\bibinfo {volume} {9}},\ \bibinfo {pages}
  {18467} (\bibinfo {year} {2019})}\BibitemShut {NoStop}%
\bibitem [{\citenamefont {Schiek}\ and\ \citenamefont
  {Baronio}(2019)}]{Schiek2019}%
  \BibitemOpen
  \bibfield  {author} {\bibinfo {author} {\bibfnamefont {R.}~\bibnamefont
  {Schiek}}\ and\ \bibinfo {author} {\bibfnamefont {F.}~\bibnamefont
  {Baronio}},\ }\href {\doibase 10.1103/PhysRevResearch.1.032036} {\bibfield
  {journal} {\bibinfo  {journal} {Phys. Rev. Research}\ }\textbf {\bibinfo
  {volume} {1}},\ \bibinfo {pages} {032036} (\bibinfo {year}
  {2019})}\BibitemShut {NoStop}%
\bibitem [{\citenamefont {Nielsen}\ \emph {et~al.}(2021)\citenamefont
  {Nielsen}, \citenamefont {Xu}, \citenamefont {Todd}, \citenamefont {Ferr\'e},
  \citenamefont {Clerc}, \citenamefont {Coen}, \citenamefont {Murdoch},\ and\
  \citenamefont {Erkintalo}}]{Nielsen2021}%
  \BibitemOpen
  \bibfield  {author} {\bibinfo {author} {\bibfnamefont {A.~U.}\ \bibnamefont
  {Nielsen}}, \bibinfo {author} {\bibfnamefont {Y.}~\bibnamefont {Xu}},
  \bibinfo {author} {\bibfnamefont {C.}~\bibnamefont {Todd}}, \bibinfo {author}
  {\bibfnamefont {M.}~\bibnamefont {Ferr\'e}}, \bibinfo {author} {\bibfnamefont
  {M.~G.}\ \bibnamefont {Clerc}}, \bibinfo {author} {\bibfnamefont
  {S.}~\bibnamefont {Coen}}, \bibinfo {author} {\bibfnamefont {S.~G.}\
  \bibnamefont {Murdoch}}, \ and\ \bibinfo {author} {\bibfnamefont
  {M.}~\bibnamefont {Erkintalo}},\ }\href {\doibase
  10.1103/PhysRevLett.127.123901} {\bibfield  {journal} {\bibinfo  {journal}
  {Phys. Rev. Lett.}\ }\textbf {\bibinfo {volume} {127}},\ \bibinfo {pages}
  {123901} (\bibinfo {year} {2021})}\BibitemShut {NoStop}%
\bibitem [{\citenamefont {Vanderhaegen}\ \emph
  {et~al.}(2021{\natexlab{a}})\citenamefont {Vanderhaegen}, \citenamefont
  {Naveau}, \citenamefont {Szriftgiser}, \citenamefont {Kudlinski},
  \citenamefont {Conforti}, \citenamefont {Mussot}, \citenamefont {Onorato},
  \citenamefont {Trillo}, \citenamefont {Chabchoub},\ and\ \citenamefont
  {Akhmediev}}]{Vanderhaegen2021PNAS}%
  \BibitemOpen
  \bibfield  {author} {\bibinfo {author} {\bibfnamefont {G.}~\bibnamefont
  {Vanderhaegen}}, \bibinfo {author} {\bibfnamefont {C.}~\bibnamefont
  {Naveau}}, \bibinfo {author} {\bibfnamefont {P.}~\bibnamefont {Szriftgiser}},
  \bibinfo {author} {\bibfnamefont {A.}~\bibnamefont {Kudlinski}}, \bibinfo
  {author} {\bibfnamefont {M.}~\bibnamefont {Conforti}}, \bibinfo {author}
  {\bibfnamefont {A.}~\bibnamefont {Mussot}}, \bibinfo {author} {\bibfnamefont
  {M.}~\bibnamefont {Onorato}}, \bibinfo {author} {\bibfnamefont
  {S.}~\bibnamefont {Trillo}}, \bibinfo {author} {\bibfnamefont
  {A.}~\bibnamefont {Chabchoub}}, \ and\ \bibinfo {author} {\bibfnamefont
  {N.}~\bibnamefont {Akhmediev}},\ }\href {\doibase 10.1073/pnas.2019348118}
  {\bibfield  {journal} {\bibinfo  {journal} {PNAS}\ }\textbf {\bibinfo
  {volume} {118}} (\bibinfo {year} {2021}{\natexlab{a}}),\
  10.1073/pnas.2019348118}\BibitemShut {NoStop}%
\bibitem [{\citenamefont {Chabchoub}\ \emph {et~al.}(2011)\citenamefont
  {Chabchoub}, \citenamefont {Hoffmann},\ and\ \citenamefont
  {Akhmediev}}]{Chabchoub2011}%
  \BibitemOpen
  \bibfield  {author} {\bibinfo {author} {\bibfnamefont {A.}~\bibnamefont
  {Chabchoub}}, \bibinfo {author} {\bibfnamefont {N.~P.}\ \bibnamefont
  {Hoffmann}}, \ and\ \bibinfo {author} {\bibfnamefont {N.}~\bibnamefont
  {Akhmediev}},\ }\href {\doibase 10.1103/PhysRevLett.106.204502} {\bibfield
  {journal} {\bibinfo  {journal} {Phys. Rev. Lett.}\ }\textbf {\bibinfo
  {volume} {106}},\ \bibinfo {pages} {204502} (\bibinfo {year}
  {2011})}\BibitemShut {NoStop}%
\bibitem [{\citenamefont {Kimmoun}\ \emph {et~al.}(2016)\citenamefont
  {Kimmoun}, \citenamefont {Hsu}, \citenamefont {Branger}, \citenamefont {Li},
  \citenamefont {Chen}, \citenamefont {Kharif}, \citenamefont {Onorato},
  \citenamefont {Kelleher}, \citenamefont {Kibler}, \citenamefont {Akhmediev},\
  and\ \citenamefont {Chabchoub}}]{Kimmoun2016}%
  \BibitemOpen
  \bibfield  {author} {\bibinfo {author} {\bibfnamefont {O.}~\bibnamefont
  {Kimmoun}}, \bibinfo {author} {\bibfnamefont {H.~C.}\ \bibnamefont {Hsu}},
  \bibinfo {author} {\bibfnamefont {H.}~\bibnamefont {Branger}}, \bibinfo
  {author} {\bibfnamefont {M.~S.}\ \bibnamefont {Li}}, \bibinfo {author}
  {\bibfnamefont {Y.~Y.}\ \bibnamefont {Chen}}, \bibinfo {author}
  {\bibfnamefont {C.}~\bibnamefont {Kharif}}, \bibinfo {author} {\bibfnamefont
  {M.}~\bibnamefont {Onorato}}, \bibinfo {author} {\bibfnamefont {E.~J.~R.}\
  \bibnamefont {Kelleher}}, \bibinfo {author} {\bibfnamefont {B.}~\bibnamefont
  {Kibler}}, \bibinfo {author} {\bibfnamefont {N.}~\bibnamefont {Akhmediev}}, \
  and\ \bibinfo {author} {\bibfnamefont {A.}~\bibnamefont {Chabchoub}},\ }\href
  {\doibase 10.1038/srep28516} {\bibfield  {journal} {\bibinfo  {journal} {Sci.
  Rep.}\ }\textbf {\bibinfo {volume} {6}},\ \bibinfo {pages} {28516} (\bibinfo
  {year} {2016})}\BibitemShut {NoStop}%
\bibitem [{\citenamefont {Eeltink}\ \emph {et~al.}(2020)\citenamefont
  {Eeltink}, \citenamefont {Armaroli}, \citenamefont {Luneau}, \citenamefont
  {Branger}, \citenamefont {Brunetti},\ and\ \citenamefont
  {Kasparian}}]{Eeltink2020}%
  \BibitemOpen
  \bibfield  {author} {\bibinfo {author} {\bibfnamefont {D.}~\bibnamefont
  {Eeltink}}, \bibinfo {author} {\bibfnamefont {A.}~\bibnamefont {Armaroli}},
  \bibinfo {author} {\bibfnamefont {C.}~\bibnamefont {Luneau}}, \bibinfo
  {author} {\bibfnamefont {H.}~\bibnamefont {Branger}}, \bibinfo {author}
  {\bibfnamefont {M.}~\bibnamefont {Brunetti}}, \ and\ \bibinfo {author}
  {\bibfnamefont {J.}~\bibnamefont {Kasparian}},\ }\href {\doibase
  https://doi.org/10.1007/s11071-020-06043-1} {\bibfield  {journal} {\bibinfo
  {journal} {Nonlinear Dyn.}\ }\textbf {\bibinfo {volume} {102}},\ \bibinfo
  {pages} {2385} (\bibinfo {year} {2020})}\BibitemShut {NoStop}%
\bibitem [{\citenamefont {Bonnefoy}\ \emph {et~al.}(2020)\citenamefont
  {Bonnefoy}, \citenamefont {Tikan}, \citenamefont {Copie}, \citenamefont
  {Suret}, \citenamefont {Ducrozet}, \citenamefont {Prabhudesai}, \citenamefont
  {Michel}, \citenamefont {Cazaubiel}, \citenamefont {Falcon}, \citenamefont
  {El},\ and\ \citenamefont {Randoux}}]{Bonnefoy2020}%
  \BibitemOpen
  \bibfield  {author} {\bibinfo {author} {\bibfnamefont {F.}~\bibnamefont
  {Bonnefoy}}, \bibinfo {author} {\bibfnamefont {A.}~\bibnamefont {Tikan}},
  \bibinfo {author} {\bibfnamefont {F.}~\bibnamefont {Copie}}, \bibinfo
  {author} {\bibfnamefont {P.}~\bibnamefont {Suret}}, \bibinfo {author}
  {\bibfnamefont {G.}~\bibnamefont {Ducrozet}}, \bibinfo {author}
  {\bibfnamefont {G.}~\bibnamefont {Prabhudesai}}, \bibinfo {author}
  {\bibfnamefont {G.}~\bibnamefont {Michel}}, \bibinfo {author} {\bibfnamefont
  {A.}~\bibnamefont {Cazaubiel}}, \bibinfo {author} {\bibfnamefont
  {E.}~\bibnamefont {Falcon}}, \bibinfo {author} {\bibfnamefont
  {G.}~\bibnamefont {El}}, \ and\ \bibinfo {author} {\bibfnamefont
  {S.}~\bibnamefont {Randoux}},\ }\href {\doibase
  10.1103/PhysRevFluids.5.034802} {\bibfield  {journal} {\bibinfo  {journal}
  {Phys. Rev. Fluids}\ }\textbf {\bibinfo {volume} {5}},\ \bibinfo {pages}
  {034802} (\bibinfo {year} {2020})}\BibitemShut {NoStop}%
\bibitem [{\citenamefont {Coppini}\ \emph {et~al.}(2020)\citenamefont
  {Coppini}, \citenamefont {Grinevich},\ and\ \citenamefont
  {Santini}}]{Coppini2020}%
  \BibitemOpen
  \bibfield  {author} {\bibinfo {author} {\bibfnamefont {F.}~\bibnamefont
  {Coppini}}, \bibinfo {author} {\bibfnamefont {P.~G.}\ \bibnamefont
  {Grinevich}}, \ and\ \bibinfo {author} {\bibfnamefont {P.~M.}\ \bibnamefont
  {Santini}},\ }\href {\doibase 10.1103/PhysRevE.101.032204} {\bibfield
  {journal} {\bibinfo  {journal} {Phys. Rev. E}\ }\textbf {\bibinfo {volume}
  {101}},\ \bibinfo {pages} {032204} (\bibinfo {year} {2020})}\BibitemShut
  {NoStop}%
\bibitem [{\citenamefont {Schober}\ and\ \citenamefont
  {Islas}(2021)}]{Schober2021}%
  \BibitemOpen
  \bibfield  {author} {\bibinfo {author} {\bibfnamefont {C.~M.}\ \bibnamefont
  {Schober}}\ and\ \bibinfo {author} {\bibfnamefont {A.~L.}\ \bibnamefont
  {Islas}},\ }\href {\doibase 10.3389/fphy.2021.633890} {\bibfield  {journal}
  {\bibinfo  {journal} {Frontiers in Physics}\ }\textbf {\bibinfo {volume}
  {9}},\ \bibinfo {pages} {95} (\bibinfo {year} {2021})}\BibitemShut {NoStop}%
\bibitem [{\citenamefont {Agrawal}(2007)}]{Agrawalbook}%
  \BibitemOpen
  \bibfield  {author} {\bibinfo {author} {\bibfnamefont {G.}~\bibnamefont
  {Agrawal}},\ }\href@noop {} {\emph {\bibinfo {title} {Nonlinear Fiber
  Optics}}}\ (\bibinfo  {publisher} {Academic Press},\ \bibinfo {year}
  {2007})\BibitemShut {NoStop}%
\bibitem [{\citenamefont {Ania-Casta\~n\'on}\ \emph {et~al.}(2006)\citenamefont
  {Ania-Casta\~n\'on}, \citenamefont {Ellingham}, \citenamefont {Ibbotson},
  \citenamefont {Chen}, \citenamefont {Zhang},\ and\ \citenamefont
  {Turitsyn}}]{AniaCastanon2006}%
  \BibitemOpen
  \bibfield  {author} {\bibinfo {author} {\bibfnamefont {J.~D.}\ \bibnamefont
  {Ania-Casta\~n\'on}}, \bibinfo {author} {\bibfnamefont {T.~J.}\ \bibnamefont
  {Ellingham}}, \bibinfo {author} {\bibfnamefont {R.}~\bibnamefont {Ibbotson}},
  \bibinfo {author} {\bibfnamefont {X.}~\bibnamefont {Chen}}, \bibinfo {author}
  {\bibfnamefont {L.}~\bibnamefont {Zhang}}, \ and\ \bibinfo {author}
  {\bibfnamefont {S.~K.}\ \bibnamefont {Turitsyn}},\ }\href {\doibase
  10.1103/PhysRevLett.96.023902} {\bibfield  {journal} {\bibinfo  {journal}
  {Phys. Rev. Lett.}\ }\textbf {\bibinfo {volume} {96}},\ \bibinfo {pages}
  {023902} (\bibinfo {year} {2006})}\BibitemShut {NoStop}%
\bibitem [{\citenamefont {Naveau}\ \emph {et~al.}(2021)\citenamefont {Naveau},
  \citenamefont {Vanderhaegen}, \citenamefont {Szriftgiser}, \citenamefont
  {Martinelli}, \citenamefont {Droques}, \citenamefont {Kudlinski},
  \citenamefont {Conforti}, \citenamefont {Trillo}, \citenamefont {Akhmediev},\
  and\ \citenamefont {Mussot}}]{Naveau2021}%
  \BibitemOpen
  \bibfield  {author} {\bibinfo {author} {\bibfnamefont {C.}~\bibnamefont
  {Naveau}}, \bibinfo {author} {\bibfnamefont {G.}~\bibnamefont
  {Vanderhaegen}}, \bibinfo {author} {\bibfnamefont {P.}~\bibnamefont
  {Szriftgiser}}, \bibinfo {author} {\bibfnamefont {G.}~\bibnamefont
  {Martinelli}}, \bibinfo {author} {\bibfnamefont {M.}~\bibnamefont {Droques}},
  \bibinfo {author} {\bibfnamefont {A.}~\bibnamefont {Kudlinski}}, \bibinfo
  {author} {\bibfnamefont {M.}~\bibnamefont {Conforti}}, \bibinfo {author}
  {\bibfnamefont {S.}~\bibnamefont {Trillo}}, \bibinfo {author} {\bibfnamefont
  {N.}~\bibnamefont {Akhmediev}}, \ and\ \bibinfo {author} {\bibfnamefont
  {A.}~\bibnamefont {Mussot}},\ }\href {\doibase 10.3389/fphy.2021.637812}
  {\bibfield  {journal} {\bibinfo  {journal} {Frontiers in Physics}\ }\textbf
  {\bibinfo {volume} {9}},\ \bibinfo {pages} {25} (\bibinfo {year}
  {2021})}\BibitemShut {NoStop}%
\bibitem [{\citenamefont {Segur}\ and\ \citenamefont
  {Henderson}(2007)}]{Segur2007}%
  \BibitemOpen
  \bibfield  {author} {\bibinfo {author} {\bibfnamefont {H.}~\bibnamefont
  {Segur}}\ and\ \bibinfo {author} {\bibfnamefont {D.~M.}\ \bibnamefont
  {Henderson}},\ }\href {\doibase 10.1140/epjst/e2007-00201-1} {\bibfield
  {journal} {\bibinfo  {journal} {Eur. Phys. J. Spec. Top.}\ }\textbf {\bibinfo
  {volume} {147}},\ \bibinfo {pages} {25} (\bibinfo {year} {2007})}\BibitemShut
  {NoStop}%
\bibitem [{\citenamefont {Akhmediev}\ and\ \citenamefont
  {Korneev}(1986)}]{Akhmediev1986}%
  \BibitemOpen
  \bibfield  {author} {\bibinfo {author} {\bibfnamefont {N.}~\bibnamefont
  {Akhmediev}}\ and\ \bibinfo {author} {\bibfnamefont {V.~I.}\ \bibnamefont
  {Korneev}},\ }\href {https://doi.org/10.1007/BF01037866} {\bibfield
  {journal} {\bibinfo  {journal} {Theor. Math. Phys.}\ }\textbf {\bibinfo
  {volume} {69}},\ \bibinfo {pages} {1089} (\bibinfo {year}
  {1986})}\BibitemShut {NoStop}%
\bibitem [{\citenamefont {Akhmediev}\ \emph {et~al.}(1987)\citenamefont
  {Akhmediev}, \citenamefont {Eleonskii},\ and\ \citenamefont
  {Kulagin}}]{Akhmediev1987}%
  \BibitemOpen
  \bibfield  {author} {\bibinfo {author} {\bibfnamefont {N.}~\bibnamefont
  {Akhmediev}}, \bibinfo {author} {\bibfnamefont {V.~M.}\ \bibnamefont
  {Eleonskii}}, \ and\ \bibinfo {author} {\bibfnamefont {N.~E.}\ \bibnamefont
  {Kulagin}},\ }\href {https://doi.org/10.1007/BF01017105} {\bibfield
  {journal} {\bibinfo  {journal} {Theor. Math. Phys.}\ }\textbf {\bibinfo
  {volume} {72}},\ \bibinfo {pages} {809} (\bibinfo {year} {1987})}\BibitemShut
  {NoStop}%
\bibitem [{\citenamefont {Trillo}\ and\ \citenamefont
  {Wabnitz}(1991{\natexlab{a}})}]{Trillo1991a}%
  \BibitemOpen
  \bibfield  {author} {\bibinfo {author} {\bibfnamefont {S.}~\bibnamefont
  {Trillo}}\ and\ \bibinfo {author} {\bibfnamefont {S.}~\bibnamefont
  {Wabnitz}},\ }\href {\doibase 10.1364/OL.16.000986} {\bibfield  {journal}
  {\bibinfo  {journal} {Opt. Lett.}\ }\textbf {\bibinfo {volume} {16}},\
  \bibinfo {pages} {986} (\bibinfo {year} {1991}{\natexlab{a}})}\BibitemShut
  {NoStop}%
\bibitem [{SM()}]{SM}%
  \BibitemOpen
  \href@noop {} {\enquote {\bibinfo {title} {See supplemental material at [url
  will be inserted by publisher] for the derivation and details of the
  perturbative results eqs. (3-4), additional measurements, and a comparison
  with damped oscillations of a pendulum.}}\ }\BibitemShut {NoStop}%
\bibitem [{\citenamefont {Vanderhaegen}\ \emph
  {et~al.}(2020{\natexlab{a}})\citenamefont {Vanderhaegen}, \citenamefont
  {Szriftgiser}, \citenamefont {Kudlinski}, \citenamefont {Conforti},
  \citenamefont {Trillo}, \citenamefont {Droques},\ and\ \citenamefont
  {Mussot}}]{Vanderhaegen2020OE}%
  \BibitemOpen
  \bibfield  {author} {\bibinfo {author} {\bibfnamefont {G.}~\bibnamefont
  {Vanderhaegen}}, \bibinfo {author} {\bibfnamefont {P.}~\bibnamefont
  {Szriftgiser}}, \bibinfo {author} {\bibfnamefont {A.}~\bibnamefont
  {Kudlinski}}, \bibinfo {author} {\bibfnamefont {M.}~\bibnamefont {Conforti}},
  \bibinfo {author} {\bibfnamefont {S.}~\bibnamefont {Trillo}}, \bibinfo
  {author} {\bibfnamefont {M.}~\bibnamefont {Droques}}, \ and\ \bibinfo
  {author} {\bibfnamefont {A.}~\bibnamefont {Mussot}},\ }\href {\doibase
  10.1364/OE.391560} {\bibfield  {journal} {\bibinfo  {journal} {Optics
  Express}\ }\textbf {\bibinfo {volume} {28}},\ \bibinfo {pages} {17773}
  (\bibinfo {year} {2020}{\natexlab{a}})}\BibitemShut {NoStop}%
\bibitem [{\citenamefont {Vanderhaegen}\ \emph
  {et~al.}(2020{\natexlab{b}})\citenamefont {Vanderhaegen}, \citenamefont
  {Szriftgiser}, \citenamefont {Naveau}, \citenamefont {Kudlinski},
  \citenamefont {Conforti}, \citenamefont {Trillo}, \citenamefont {Akhmediev},\
  and\ \citenamefont {Mussot}}]{Vanderhaegen2020OL}%
  \BibitemOpen
  \bibfield  {author} {\bibinfo {author} {\bibfnamefont {G.}~\bibnamefont
  {Vanderhaegen}}, \bibinfo {author} {\bibfnamefont {P.}~\bibnamefont
  {Szriftgiser}}, \bibinfo {author} {\bibfnamefont {C.}~\bibnamefont {Naveau}},
  \bibinfo {author} {\bibfnamefont {A.}~\bibnamefont {Kudlinski}}, \bibinfo
  {author} {\bibfnamefont {M.}~\bibnamefont {Conforti}}, \bibinfo {author}
  {\bibfnamefont {S.}~\bibnamefont {Trillo}}, \bibinfo {author} {\bibfnamefont
  {N.}~\bibnamefont {Akhmediev}}, \ and\ \bibinfo {author} {\bibfnamefont
  {A.}~\bibnamefont {Mussot}},\ }\href {\doibase 10.1364/OL.394604} {\bibfield
  {journal} {\bibinfo  {journal} {Opt. Lett.}\ }\textbf {\bibinfo {volume}
  {45}},\ \bibinfo {pages} {3757} (\bibinfo {year}
  {2020}{\natexlab{b}})}\BibitemShut {NoStop}%
\bibitem [{\citenamefont {Vanderhaegen}\ \emph
  {et~al.}(2021{\natexlab{b}})\citenamefont {Vanderhaegen}, \citenamefont
  {Szriftgiser}, \citenamefont {Conforti}, \citenamefont {Kudlinski},
  \citenamefont {Droques},\ and\ \citenamefont {Mussot}}]{Vanderhaegen2021b}%
  \BibitemOpen
  \bibfield  {author} {\bibinfo {author} {\bibfnamefont {G.}~\bibnamefont
  {Vanderhaegen}}, \bibinfo {author} {\bibfnamefont {P.}~\bibnamefont
  {Szriftgiser}}, \bibinfo {author} {\bibfnamefont {M.}~\bibnamefont
  {Conforti}}, \bibinfo {author} {\bibfnamefont {A.}~\bibnamefont {Kudlinski}},
  \bibinfo {author} {\bibfnamefont {M.}~\bibnamefont {Droques}}, \ and\
  \bibinfo {author} {\bibfnamefont {A.}~\bibnamefont {Mussot}},\ }\href
  {\doibase 10.1364/OL.434956} {\bibfield  {journal} {\bibinfo  {journal}
  {Optics Letters}\ }\textbf {\bibinfo {volume} {46}},\ \bibinfo {pages} {5019}
  (\bibinfo {year} {2021}{\natexlab{b}})}\BibitemShut {NoStop}%
\bibitem [{\citenamefont {Ercolani}\ \emph {et~al.}(1990)\citenamefont
  {Ercolani}, \citenamefont {Forest},\ and\ \citenamefont
  {McLaughlin}}]{Ercolani1990_SGeq}%
  \BibitemOpen
  \bibfield  {author} {\bibinfo {author} {\bibfnamefont {N.}~\bibnamefont
  {Ercolani}}, \bibinfo {author} {\bibfnamefont {M.}~\bibnamefont {Forest}}, \
  and\ \bibinfo {author} {\bibfnamefont {D.~W.}\ \bibnamefont {McLaughlin}},\
  }\href {\doibase https://doi.org/10.1016/0167-2789(90)90142-C} {\bibfield
  {journal} {\bibinfo  {journal} {Physica D}\ }\textbf {\bibinfo {volume}
  {43}},\ \bibinfo {pages} {349} (\bibinfo {year} {1990})}\BibitemShut
  {NoStop}%
\bibitem [{\citenamefont {Chen}\ \emph {et~al.}(2021)\citenamefont {Chen},
  \citenamefont {Liu}, \citenamefont {Yao}, \citenamefont {Zhao},\ and\
  \citenamefont {Akhmediev}}]{Chen2020_Manakov}%
  \BibitemOpen
  \bibfield  {author} {\bibinfo {author} {\bibfnamefont {S.-C.}\ \bibnamefont
  {Chen}}, \bibinfo {author} {\bibfnamefont {C.}~\bibnamefont {Liu}}, \bibinfo
  {author} {\bibfnamefont {X.}~\bibnamefont {Yao}}, \bibinfo {author}
  {\bibfnamefont {L.-C.}\ \bibnamefont {Zhao}}, \ and\ \bibinfo {author}
  {\bibfnamefont {N.}~\bibnamefont {Akhmediev}},\ }\href {\doibase
  10.1103/PhysRevE.104.024215} {\bibfield  {journal} {\bibinfo  {journal}
  {Phys. Rev. E}\ }\textbf {\bibinfo {volume} {104}},\ \bibinfo {pages}
  {024215} (\bibinfo {year} {2021})}\BibitemShut {NoStop}%
\bibitem [{\citenamefont {Liu}\ \emph {et~al.}(2021)\citenamefont {Liu},
  \citenamefont {Wu}, \citenamefont {Chen}, \citenamefont {Yao},\ and\
  \citenamefont {Akhmediev}}]{Liu2021_CLLeq}%
  \BibitemOpen
  \bibfield  {author} {\bibinfo {author} {\bibfnamefont {C.}~\bibnamefont
  {Liu}}, \bibinfo {author} {\bibfnamefont {Y.-H.}\ \bibnamefont {Wu}},
  \bibinfo {author} {\bibfnamefont {S.-C.}\ \bibnamefont {Chen}}, \bibinfo
  {author} {\bibfnamefont {X.}~\bibnamefont {Yao}}, \ and\ \bibinfo {author}
  {\bibfnamefont {N.}~\bibnamefont {Akhmediev}},\ }\href {\doibase
  10.1103/PhysRevLett.127.094102} {\bibfield  {journal} {\bibinfo  {journal}
  {Phys. Rev. Lett.}\ }\textbf {\bibinfo {volume} {127}},\ \bibinfo {pages}
  {094102} (\bibinfo {year} {2021})}\BibitemShut {NoStop}%
\bibitem [{\citenamefont {Trillo}\ and\ \citenamefont
  {Wabnitz}(1991{\natexlab{b}})}]{Trillo1991_VNLS}%
  \BibitemOpen
  \bibfield  {author} {\bibinfo {author} {\bibfnamefont {S.}~\bibnamefont
  {Trillo}}\ and\ \bibinfo {author} {\bibfnamefont {S.}~\bibnamefont
  {Wabnitz}},\ }\href {\doibase https://doi.org/10.1016/0375-9601(91)90519-E}
  {\bibfield  {journal} {\bibinfo  {journal} {Phys. Lett. A}\ }\textbf
  {\bibinfo {volume} {159}},\ \bibinfo {pages} {252} (\bibinfo {year}
  {1991}{\natexlab{b}})}\BibitemShut {NoStop}%
\bibitem [{\citenamefont {Conforti}\ \emph {et~al.}(2016)\citenamefont
  {Conforti}, \citenamefont {Mussot}, \citenamefont {Kudlinski}, \citenamefont
  {Rota~Nodari}, \citenamefont {Dujardin}, \citenamefont {De~Bi\'evre},
  \citenamefont {Armaroli},\ and\ \citenamefont {Trillo}}]{Conforti2016}%
  \BibitemOpen
  \bibfield  {author} {\bibinfo {author} {\bibfnamefont {M.}~\bibnamefont
  {Conforti}}, \bibinfo {author} {\bibfnamefont {A.}~\bibnamefont {Mussot}},
  \bibinfo {author} {\bibfnamefont {A.}~\bibnamefont {Kudlinski}}, \bibinfo
  {author} {\bibfnamefont {S.}~\bibnamefont {Rota~Nodari}}, \bibinfo {author}
  {\bibfnamefont {G.}~\bibnamefont {Dujardin}}, \bibinfo {author}
  {\bibfnamefont {S.}~\bibnamefont {De~Bi\'evre}}, \bibinfo {author}
  {\bibfnamefont {A.}~\bibnamefont {Armaroli}}, \ and\ \bibinfo {author}
  {\bibfnamefont {S.}~\bibnamefont {Trillo}},\ }\href {\doibase
  10.1103/PhysRevLett.117.013901} {\bibfield  {journal} {\bibinfo  {journal}
  {Phys. Rev. Lett.}\ }\textbf {\bibinfo {volume} {117}},\ \bibinfo {pages}
  {013901} (\bibinfo {year} {2016})}\BibitemShut {NoStop}%
\end{thebibliography}%


\begin{thebibliography}{11}%
\makeatletter
\providecommand \@ifxundefined [1]{%
 \@ifx{#1\undefined}
}%
\providecommand \@ifnum [1]{%
 \ifnum #1\expandafter \@firstoftwo
 \else \expandafter \@secondoftwo
 \fi
}%
\providecommand \@ifx [1]{%
 \ifx #1\expandafter \@firstoftwo
 \else \expandafter \@secondoftwo
 \fi
}%
\providecommand \natexlab [1]{#1}%
\providecommand \enquote  [1]{``#1''}%
\providecommand \bibnamefont  [1]{#1}%
\providecommand \bibfnamefont [1]{#1}%
\providecommand \citenamefont [1]{#1}%
\providecommand \href@noop [0]{\@secondoftwo}%
\providecommand \href [0]{\begingroup \@sanitize@url \@href}%
\providecommand \@href[1]{\@@startlink{#1}\@@href}%
\providecommand \@@href[1]{\endgroup#1\@@endlink}%
\providecommand \@sanitize@url [0]{\catcode `\\12\catcode `\$12\catcode
  `\&12\catcode `\#12\catcode `\^12\catcode `\_12\catcode `\%12\relax}%
\providecommand \@@startlink[1]{}%
\providecommand \@@endlink[0]{}%
\providecommand \url  [0]{\begingroup\@sanitize@url \@url }%
\providecommand \@url [1]{\endgroup\@href {#1}{\urlprefix }}%
\providecommand \urlprefix  [0]{URL }%
\providecommand \Eprint [0]{\href }%
\providecommand \doibase [0]{http://dx.doi.org/}%
\providecommand \selectlanguage [0]{\@gobble}%
\providecommand \bibinfo  [0]{\@secondoftwo}%
\providecommand \bibfield  [0]{\@secondoftwo}%
\providecommand \translation [1]{[#1]}%
\providecommand \BibitemOpen [0]{}%
\providecommand \bibitemStop [0]{}%
\providecommand \bibitemNoStop [0]{.\EOS\space}%
\providecommand \EOS [0]{\spacefactor3000\relax}%
\providecommand \BibitemShut  [1]{\csname bibitem#1\endcsname}%
\let\auto@bib@innerbib\@empty
\bibitem [{\citenamefont {Trillo}\ and\ \citenamefont
  {Wabnitz}(1991)}]{Trillo1991a}%
  \BibitemOpen
  \bibfield  {author} {\bibinfo {author} {\bibfnamefont {S.}~\bibnamefont
  {Trillo}}\ and\ \bibinfo {author} {\bibfnamefont {S.}~\bibnamefont
  {Wabnitz}},\ }\href {\doibase 10.1364/OL.16.000986} {\bibfield  {journal}
  {\bibinfo  {journal} {Opt. Lett.}\ }\textbf {\bibinfo {volume} {16}},\
  \bibinfo {pages} {986} (\bibinfo {year} {1991})}\BibitemShut {NoStop}%
\bibitem [{\citenamefont {Cappellini}\ and\ \citenamefont
  {Trillo}(1991)}]{Cappellini1991}%
  \BibitemOpen
  \bibfield  {author} {\bibinfo {author} {\bibfnamefont {G.}~\bibnamefont
  {Cappellini}}\ and\ \bibinfo {author} {\bibfnamefont {S.}~\bibnamefont
  {Trillo}},\ }\href {\doibase 10.1364/JOSAB.8.000824} {\bibfield  {journal}
  {\bibinfo  {journal} {Journal of the Optical Society of America B}\ }\textbf
  {\bibinfo {volume} {8}},\ \bibinfo {pages} {824} (\bibinfo {year}
  {1991})}\BibitemShut {NoStop}%
\bibitem [{\citenamefont {Mussot}\ \emph {et~al.}(2018)\citenamefont {Mussot},
  \citenamefont {Naveau}, \citenamefont {Conforti}, \citenamefont {Kudlinski},
  \citenamefont {Copie}, \citenamefont {Szriftgiser},\ and\ \citenamefont
  {Trillo}}]{Mussot2018}%
  \BibitemOpen
  \bibfield  {author} {\bibinfo {author} {\bibfnamefont {A.}~\bibnamefont
  {Mussot}}, \bibinfo {author} {\bibfnamefont {C.}~\bibnamefont {Naveau}},
  \bibinfo {author} {\bibfnamefont {M.}~\bibnamefont {Conforti}}, \bibinfo
  {author} {\bibfnamefont {A.}~\bibnamefont {Kudlinski}}, \bibinfo {author}
  {\bibfnamefont {F.}~\bibnamefont {Copie}}, \bibinfo {author} {\bibfnamefont
  {P.}~\bibnamefont {Szriftgiser}}, \ and\ \bibinfo {author} {\bibfnamefont
  {S.}~\bibnamefont {Trillo}},\ }\href {\doibase 10.1038/s41566-018-0136-1}
  {\bibfield  {journal} {\bibinfo  {journal} {Nat. Photonics}\ }\textbf
  {\bibinfo {volume} {12}},\ \bibinfo {pages} {303} (\bibinfo {year}
  {2018})}\BibitemShut {NoStop}%
\bibitem [{\citenamefont {Coppini}\ \emph {et~al.}(2020)\citenamefont
  {Coppini}, \citenamefont {Grinevich},\ and\ \citenamefont
  {Santini}}]{Coppini2020}%
  \BibitemOpen
  \bibfield  {author} {\bibinfo {author} {\bibfnamefont {F.}~\bibnamefont
  {Coppini}}, \bibinfo {author} {\bibfnamefont {P.~G.}\ \bibnamefont
  {Grinevich}}, \ and\ \bibinfo {author} {\bibfnamefont {P.~M.}\ \bibnamefont
  {Santini}},\ }\href {\doibase 10.1103/PhysRevE.101.032204} {\bibfield
  {journal} {\bibinfo  {journal} {Phys. Rev. E}\ }\textbf {\bibinfo {volume}
  {101}},\ \bibinfo {pages} {032204} (\bibinfo {year} {2020})}\BibitemShut
  {NoStop}%
\bibitem [{\citenamefont {Grinevich}\ and\ \citenamefont
  {Santini}(2018)}]{GS2018a}%
  \BibitemOpen
  \bibfield  {author} {\bibinfo {author} {\bibfnamefont {P.}~\bibnamefont
  {Grinevich}}\ and\ \bibinfo {author} {\bibfnamefont {P.}~\bibnamefont
  {Santini}},\ }\href {\doibase https://doi.org/10.1016/j.physleta.2018.02.014}
  {\bibfield  {journal} {\bibinfo  {journal} {Phys. Lett. A}\ }\textbf
  {\bibinfo {volume} {382}},\ \bibinfo {pages} {973} (\bibinfo {year}
  {2018})}\BibitemShut {NoStop}%
\bibitem [{\citenamefont {Trillo}\ and\ \citenamefont
  {Conforti}(2019)}]{Trillo2019}%
  \BibitemOpen
  \bibfield  {author} {\bibinfo {author} {\bibfnamefont {S.}~\bibnamefont
  {Trillo}}\ and\ \bibinfo {author} {\bibfnamefont {M.}~\bibnamefont
  {Conforti}},\ }\href {\doibase 10.1364/OL.44.004275} {\bibfield  {journal}
  {\bibinfo  {journal} {Opt. Lett.}\ }\textbf {\bibinfo {volume} {44}},\
  \bibinfo {pages} {4275} (\bibinfo {year} {2019})}\BibitemShut {NoStop}%
\bibitem [{\citenamefont {Pierangeli}\ \emph {et~al.}(2018)\citenamefont
  {Pierangeli}, \citenamefont {Flammini}, \citenamefont {Zhang}, \citenamefont
  {Marcucci}, \citenamefont {Agranat}, \citenamefont {Grinevich}, \citenamefont
  {Santini}, \citenamefont {Conti},\ and\ \citenamefont
  {DelRe}}]{Pierangeli2018}%
  \BibitemOpen
  \bibfield  {author} {\bibinfo {author} {\bibfnamefont {D.}~\bibnamefont
  {Pierangeli}}, \bibinfo {author} {\bibfnamefont {M.}~\bibnamefont
  {Flammini}}, \bibinfo {author} {\bibfnamefont {L.}~\bibnamefont {Zhang}},
  \bibinfo {author} {\bibfnamefont {G.}~\bibnamefont {Marcucci}}, \bibinfo
  {author} {\bibfnamefont {A.~J.}\ \bibnamefont {Agranat}}, \bibinfo {author}
  {\bibfnamefont {P.~G.}\ \bibnamefont {Grinevich}}, \bibinfo {author}
  {\bibfnamefont {P.~M.}\ \bibnamefont {Santini}}, \bibinfo {author}
  {\bibfnamefont {C.}~\bibnamefont {Conti}}, \ and\ \bibinfo {author}
  {\bibfnamefont {E.}~\bibnamefont {DelRe}},\ }\href {\doibase
  10.1103/PhysRevX.8.041017} {\bibfield  {journal} {\bibinfo  {journal} {Phys.
  Rev. X}\ }\textbf {\bibinfo {volume} {8}},\ \bibinfo {pages} {041017}
  (\bibinfo {year} {2018})}\BibitemShut {NoStop}%
\bibitem [{\citenamefont {Naveau}\ \emph {et~al.}(2019)\citenamefont {Naveau},
  \citenamefont {Szriftgiser}, \citenamefont {Kudlinski}, \citenamefont
  {Conforti}, \citenamefont {Trillo},\ and\ \citenamefont
  {Mussot}}]{Naveau2019b}%
  \BibitemOpen
  \bibfield  {author} {\bibinfo {author} {\bibfnamefont {C.}~\bibnamefont
  {Naveau}}, \bibinfo {author} {\bibfnamefont {P.}~\bibnamefont {Szriftgiser}},
  \bibinfo {author} {\bibfnamefont {A.}~\bibnamefont {Kudlinski}}, \bibinfo
  {author} {\bibfnamefont {M.}~\bibnamefont {Conforti}}, \bibinfo {author}
  {\bibfnamefont {S.}~\bibnamefont {Trillo}}, \ and\ \bibinfo {author}
  {\bibfnamefont {A.}~\bibnamefont {Mussot}},\ }\href {\doibase
  10.1364/OL.44.005426} {\bibfield  {journal} {\bibinfo  {journal} {Opt.
  Lett.}\ }\textbf {\bibinfo {volume} {44}},\ \bibinfo {pages} {5426} (\bibinfo
  {year} {2019})}\BibitemShut {NoStop}%
\bibitem [{\citenamefont {Naveau}\ \emph {et~al.}(2021)\citenamefont {Naveau},
  \citenamefont {Vanderhaegen}, \citenamefont {Szriftgiser}, \citenamefont
  {Martinelli}, \citenamefont {Droques}, \citenamefont {Kudlinski},
  \citenamefont {Conforti}, \citenamefont {Trillo}, \citenamefont {Akhmediev},\
  and\ \citenamefont {Mussot}}]{Mussot2021}%
  \BibitemOpen
  \bibfield  {author} {\bibinfo {author} {\bibfnamefont {C.}~\bibnamefont
  {Naveau}}, \bibinfo {author} {\bibfnamefont {G.}~\bibnamefont
  {Vanderhaegen}}, \bibinfo {author} {\bibfnamefont {P.}~\bibnamefont
  {Szriftgiser}}, \bibinfo {author} {\bibfnamefont {G.}~\bibnamefont
  {Martinelli}}, \bibinfo {author} {\bibfnamefont {M.}~\bibnamefont {Droques}},
  \bibinfo {author} {\bibfnamefont {A.}~\bibnamefont {Kudlinski}}, \bibinfo
  {author} {\bibfnamefont {M.}~\bibnamefont {Conforti}}, \bibinfo {author}
  {\bibfnamefont {S.}~\bibnamefont {Trillo}}, \bibinfo {author} {\bibfnamefont
  {N.}~\bibnamefont {Akhmediev}}, \ and\ \bibinfo {author} {\bibfnamefont
  {A.}~\bibnamefont {Mussot}},\ }\href {\doibase 10.3389/fphy.2021.637812}
  {\bibfield  {journal} {\bibinfo  {journal} {Front. Phys.}\ }\textbf {\bibinfo
  {volume} {9}} (\bibinfo {year} {2021}),\
  10.3389/fphy.2021.637812}\BibitemShut {NoStop}%
\bibitem [{\citenamefont {Agrawal}(2007)}]{Agrawalbook}%
  \BibitemOpen
  \bibfield  {author} {\bibinfo {author} {\bibfnamefont {G.}~\bibnamefont
  {Agrawal}},\ }\href@noop {} {\emph {\bibinfo {title} {Nonlinear Fiber
  Optics}}}\ (\bibinfo  {publisher} {Academic Press},\ \bibinfo {year}
  {2007})\BibitemShut {NoStop}%
\bibitem [{\citenamefont {Butikov}(1999)}]{Butikov1999}%
  \BibitemOpen
  \bibfield  {author} {\bibinfo {author} {\bibfnamefont {E.~I.}\ \bibnamefont
  {Butikov}},\ }\href {\doibase 10.1088/0143-0807/20/6/308} {\bibfield
  {journal} {\bibinfo  {journal} {Eur. J. Phys.}\ }\textbf {\bibinfo {volume}
  {20}},\ \bibinfo {pages} {429} (\bibinfo {year} {1999})}\BibitemShut
  {NoStop}%
\end{thebibliography}%

\end{document}


\begin{center} {\it Supplemental material to the paper:} \end{center} 
\title{
Multiple symmetry breaking induced by weak damping in the Fermi-Pasta-Ulam-Tsingou recurrence process}

\author{Guillaume Vanderhaegen}
\affiliation{University of Lille, CNRS, UMR 8523-PhLAM-Physique des Lasers Atomes et Mol\'ecules, F-59000 Lille, France}
\author{Pascal Szriftgiser}
\affiliation{University of Lille, CNRS, UMR 8523-PhLAM-Physique des Lasers Atomes et Mol\'ecules, F-59000 Lille, France}
\author{Alexandre Kudlinski}
\affiliation{University of Lille, CNRS, UMR 8523-PhLAM-Physique des Lasers Atomes et Mol\'ecules, F-59000 Lille, France}
\author{Andrea Armaroli}
\affiliation{University of Lille, CNRS, UMR 8523-PhLAM-Physique des Lasers Atomes et Mol\'ecules, F-59000 Lille, France}
\author{Matteo Conforti}
\affiliation{University of Lille, CNRS, UMR 8523-PhLAM-Physique des Lasers Atomes et Mol\'ecules, F-59000 Lille, France}
\author{Arnaud Mussot}
\affiliation{University of Lille, CNRS, UMR 8523-PhLAM-Physique des Lasers Atomes et Mol\'ecules, F-59000 Lille, France}
\author{Stefano Trillo} 
\affiliation{Department of Engineering, University of Ferrara, 44122 Ferrara, Italy}

\date{\today} 
\maketitle

\section{Perturbative approaches to predict the critical damping coefficients}
\subsection{3-wave model with damping}
The aim of this section is to show that, despite the dissipative nature of the problem, the 3-wave mixing (3WM) approach mentioned in the text can lead to a Hamiltonian formulation, though - importantly -  of the non-autonomous type.
To this end, we find convenient to start from the NLSE in dimensional units
\begin{equation} \label{NLSdim}
i\frac{\partial U}{\partial Z} - \frac{\beta_2}{2} \frac{\partial^2 U}{\partial T^2} + \gamma |U|^2 U  = -i \frac{\alpha_P}{2} U.
\end{equation}
We consider for the field $U=U(Z,T)$, the following 3-wave ansatz (for sake of simplicity, with symmetric sidebands)
\begin{equation} \label{ansatz}
\hspace{-0.5cm} U= \sqrt{P_0} \left[ a_0(Z) + \frac{a_1(Z)}{\sqrt{2}}\left(e^{i \Omega T} + e^{-i \Omega T} \right) \right] e^{-\frac{\alpha_P}{2}Z},
\end{equation}
where $P_0$ is the input power, $a_{0,1}(Z)$ are the complex amplitudes of the pump and sidebands, respectively. Since the last real exponential accounts explicitly for the damping,
$a_{0,1}(Z)$ are {\em undamped} variables, which satisfy power conservation $|a_0|^2 + |a_1|^2=1$ at any distance $Z$.
By inserting Eq. (\ref{ansatz}) into Eq. (\ref{NLSdim}), we find, neglecting generation of higher-order sideband pairs at $\pm m \Omega$, $m \ge 2$, the following coupled-equations
\begin{eqnarray}
\hspace{-0.5cm} -i\frac{\partial a_0}{\partial Z} &=& \hat{\gamma} \left[ (|a_0|^2+2|a_1|^2) a_0 + a_1^2 a_0^* \right],\label{2wavea} \\ 
\hspace{-0.5cm} -i\frac{\partial a_1}{\partial Z} &=& dk a_1 + \hat{\gamma} \left[ \left(\frac{3}{2} |a_1|^2+2|a_0|^2 \right) a_1 + a_0^2 a_{1}^* \right], \label{2waveb}
\end{eqnarray}
where $dk=\beta_2 \frac{\Omega^2}{2}$ is the dispersive mismatch, $\hat{\gamma} \equiv \hat{\gamma}(Z) = \gamma P_0~ \exp(-\alpha_P Z) = Z_{nl}^{-1} \exp(-\alpha_P Z)$ is an effective damped ($z$-dependent) nonlinear coefficient, and $Z_{nl}=(\gamma P_0)^{-1}$ stands for the characteristic nonlinear length associated with input power $P_0$.

By transforming from variables $a_{0,1}(z)$, to the pair of Hamiltonian conjugated variables $\eta(\bar{z})=|a_1(\bar{z})|^2=1-|a_0(\bar{z})|^2$ and $\Delta \phi(\bar{z})=Arg[a_1(\bar{z})]-Arg[a_0(\bar{z})]$,
and introducing the normalized effective distance $\bar{z} = \frac{1-\exp(-\alpha z)}{\alpha}$, 
where $z \equiv Z/Z_{nl}$ and $\alpha \equiv \alpha_P Z_{nl}$, we cast Eqs. (\ref{2wavea}-\ref{2waveb}) in the following Hamiltonian form
\begin{eqnarray}
\hspace{-0.5cm}  \frac{d\eta}{d\bar{z}} &=& \frac{\partial H}{\partial \Delta \phi};\;\;\;\frac{d\Delta \phi}{d\bar{z}}=-\frac{\partial H}{\partial \eta},\label{Heqs} \\
\hspace{-0.5cm} H &=& \eta (1-\eta) \cos 2 \phi + \left[1-\frac{\omega^2(\bar{z})}{2} \right] \eta -\frac{3}{4} \eta^2. \label{H}
\end{eqnarray}
Equations are of the same form of the unperturbed ($\alpha=0$) 3WM system \cite{Trillo1991a,Cappellini1991,Mussot2018}, with the fundamental difference that the system becomes non-autonomous, due to the fact that now the Hamiltonian (energy) $H=H(\eta, \phi, \bar{z})$ turns out to depend on the distance through the single dimensionless parameter $\omega^2(\bar{z})=\frac{\omega_0^2}{1-\alpha \bar{z}}$. Here the input parameter $\omega(\bar{z}=0)=\omega_0=\Omega \sqrt{|\beta_2|Z_{nl}}$ is nothing but the normalized modulation frequency which fully characterizes the unperturbed (undamped) evolutions along with the initial condition in terms of sidebands fraction and overall phase \cite{Mussot2018}. 
Importantly, in the presence of damping, the frequency parameter $\omega(\bar{z})$ increases exponentially upon propagation because the total power (at denominator) is exponentially damped. 
Furthermore, by taking the total derivative of Eq. (\ref{H}) and exploiting the Hamiltonian constraint $\frac{\partial H}{\partial \eta}  \frac{d\eta}{d\bar{z}} + \frac{\partial H}{\partial \Delta \phi} \frac{d\Delta \phi}{d\bar{z}}=0$, we obtain the equation that rules the variation of the Hamiltonian along the motion in the form
\begin{equation} \label{Hev1}
\frac{dH}{d \bar{z}}=  \frac{\partial H}{\partial \bar{z}}  = - \frac{\alpha}{2}  \frac{\omega_0^2}{(1-\alpha \bar{z})^2} \eta(\bar{z}).
\end{equation}
Equation (\ref{Hev1}) can be more conveniently rewritten by returning to the evolution variable $z = \alpha^{-1} \log(1-\alpha \bar{z} )^{-1}$, obtaining
\begin{eqnarray} \label{Hev2}
\frac{dH}{dz} =  - \frac{\alpha}{2}~\omega^2(z)~\eta(z);~~\omega^2(z) \equiv \omega_0^2e^{\alpha z}.
\end{eqnarray}
The integration of Eq. (\ref{Hev2}) between $z=0$ and $z=n z_{per}$ ($n$ periods) of the unshifted orbit, with corresponding values of the Hamiltonian $H_0 \equiv H(\eta_0,\phi_0=0,z=0)$ and $H_{n} \equiv H(n z_{per})$, respectively, yields
\begin{equation} \label{integrateH}
H_{n} - H_0 = -\frac{\alpha \omega_0^2}{2}~\int_{0}^{n z_{per}} e^{\alpha z}~\eta(z) dz,
\end{equation}
where the RHS gives the variation of the Hamiltonian $\Delta H(\alpha) \equiv - \frac{\alpha \omega_0^2}{2}~\int_{0}^{n z_{per}} e^{\alpha z}~\eta(z) dz$, which can be considered a function of  $\alpha$
for fixed initial condition and choice of $\omega_0$.
\begin{figure}[t!]
\includegraphics[width=10cm]{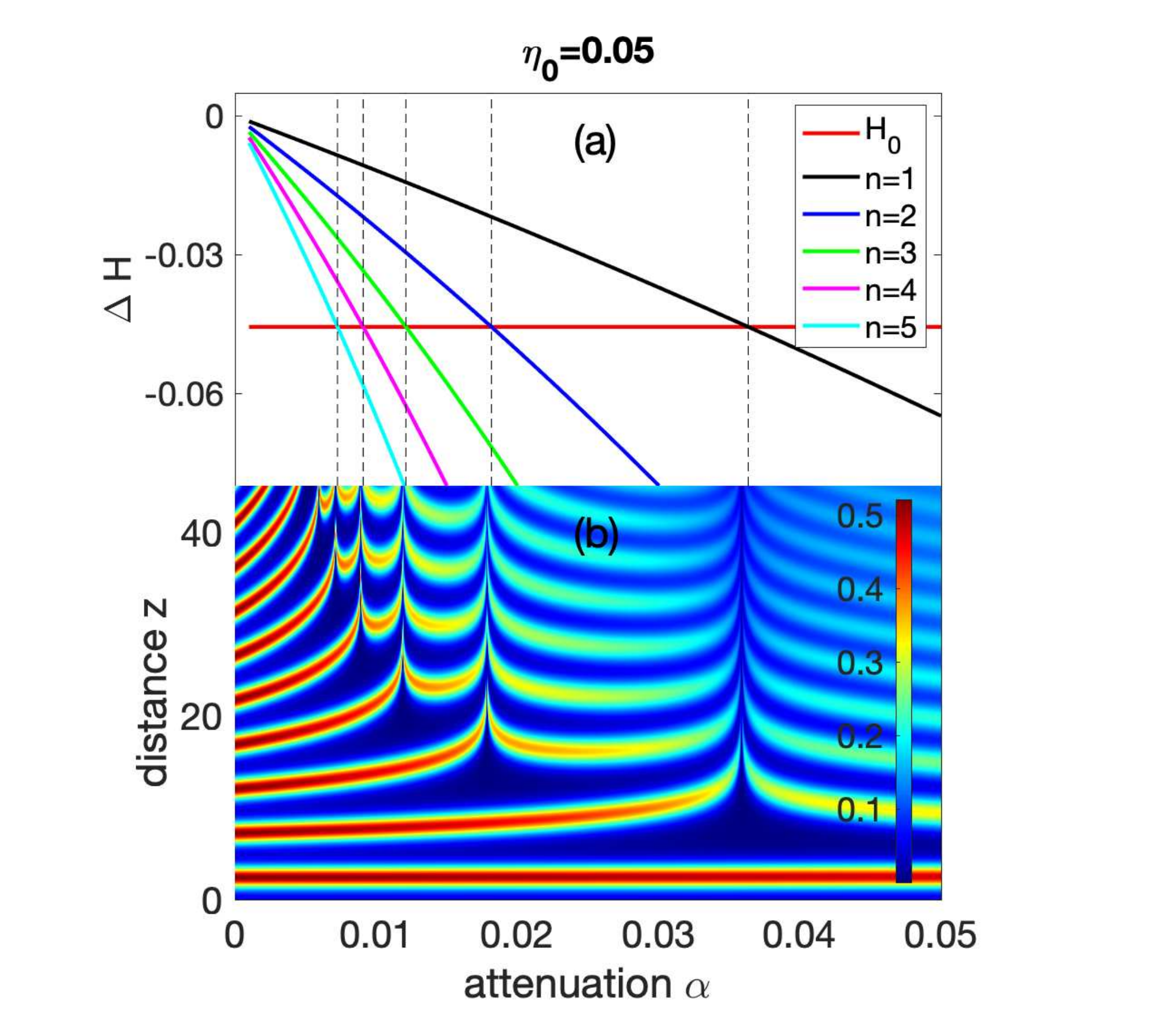}
\caption{(a) Variation of the Hamiltonian $\Delta H(\alpha)$ vs. attenuation $\alpha$, for $n=1,2,3,4,5$ passages in the unshifted orbit. Vertical dashed lines mark the critical damping values $\alpha=\alpha_{cn}$, given by crossing of $\Delta H(\alpha)$ with $-H_0$ (red horizontal line). (b) False color plot of sidebands power fraction $\eta(z)$ vs. $\alpha$ obtained from numerical integration of Eqs. (\ref{2wavea}-\ref{2waveb});
Here $\omega=\sqrt{2}$ and $\eta_0=0.05$, as in Fig. 1 of the paper.
}
\label{figS1}
\end{figure}
The $n-th$ critical value of the attenuation $\alpha$ can be estimated by assuming that the evolution is such that, after $n$ complete unshifted orbits, the point arrives, in phase plane ($\eta \cos \phi, \eta \sin \phi$), to its closest approach to the saddle point (i.e., the origin, which represents the MI-unstable pump) with the same energy pertaining to the saddle, i.e.  $H_n=H_{saddle} \equiv H(\eta=0)=0$. Using Eq. (\ref{integrateH}) with $H_n=0$, the critical attenuations satisfy Eq. (3) of the paper, which we repeat here in implicit form
\begin{equation} \label{alfac_3WM}
-H_0 = \Delta H(\alpha_{cn}).
\end{equation}
In order to verify Eq. (\ref{alfac_3WM}), we have reported in Fig. \ref{figS1}(a) the variation of the Hamiltonian $\Delta H(\alpha)$ [RHS in Eq. (\ref{integrateH})] vs. $\alpha$ for $n=1,2,3,4,5$. The integral is calculated by using the unperturbed values ($\alpha=0$) of $z_{per}$ and $\eta(z)$, due to smallness of $\alpha$. The crossing values with the constant $-H_0$, gives the critical values $\alpha=\alpha_{cn}$, marked by vertical dashed lines in Fig. \ref{figS1}(a). They turn out to be in very good agreement with the values obtained from numerical integration of the damped 3-wave equations [Eqs. (\ref{2wavea}-\ref{2waveb})], reported in Fig. \ref{figS1}(b). Therefore, this simple approach gives a clear insight on the physical mechanism of FPUT separatrix crossing which is behind the existence of the multiple critical losses. From the quantitative point of view, however, the critical values $\alpha_{cn}$ from Eq. (\ref{alfac_3WM}) are found to overestimate those arising from the NLSE, due to the fact that the 3-wave approximation overestimates the period $z_{per}$ of the recurrence.

\subsection{Perturbative finite-gap approach}

Based on a different approach, the impact of losses can be assessed by means of a perturbation theory, developed by Coppini et al. in Ref. \cite{Coppini2020}, based on the inverse scattering method for the NLSE with periodic boundary conditions, generally known as finite-gap theory of the NLSE. A convenient form of the initial condition is, in this case,
\begin{equation} \label{Sinput}
\psi_0=\sqrt{p} \left\{1+ \varepsilon \left[c_1 \exp(i \omega t) + c_{-1} \exp(-i \omega t) \right] \right\},
\end{equation}
where the perturbation theory requires small sidebands, i.e. $\varepsilon \ll 1$ ($|c_{\pm}|=O(1)$ accounts for details of the input perturbation, i.e. sideband relative phase and possible imbalance), as well as weak attenuation $\alpha \ll1$.
Without loss of generality, Eq. (\ref{Sinput}) becomes identical to the initial condition used in the paper [Eq. (2)] by posing $\varepsilon \equiv a$ and $c_1=c_{-1}=\exp(-i\phi_0)/\sqrt{2}$.

Under the hypothesis of sufficiently small sidebands and losses, one can derive, despite the complexity of the problem, the following remarkably simple leading-order expression for the distance $z_m$, $m=1,2,3, \ldots$,
at which the $m$-th peak amplification occurs in the recurrent pattern \cite{Coppini2020}
\begin{eqnarray}
z_m &=& z_0 + \sum_{n>0}^{m-1} \frac{2}{g} \log \left(\frac{g^4}{4 p^4 \varepsilon^2 |f_n(\alpha)|} \right), \label{GSloss1} \\
z_0 &=& \frac{2}{g} \log \left(\frac{g^2}{2 p^2 \varepsilon |e_+|} \right),\label{GSloss2} \\
f_n(\alpha) &=& e_+ e_- - n  \frac{\alpha}{ \varepsilon^2 p^2} g,\label{GSloss3}
\end{eqnarray}
where $g=g(\omega_0)=\omega_0\sqrt{4p-\omega_0^2}$ is the MI gain, and $e_{\pm}=\exp(\mp i\phi_{\omega}) c_{\pm 1}^*- \exp(\pm i\phi_{\omega})  c_{\mp 1}$ stand for the growing and decaying eigenvectors of the linearized stage of MI \cite{GS2018a,Trillo2019}, and $\phi_{\omega}=\cos^{-1} \left( \omega/2\sqrt{p} \right)$.
\begin{figure}[t!]
\includegraphics[width=8.6cm]{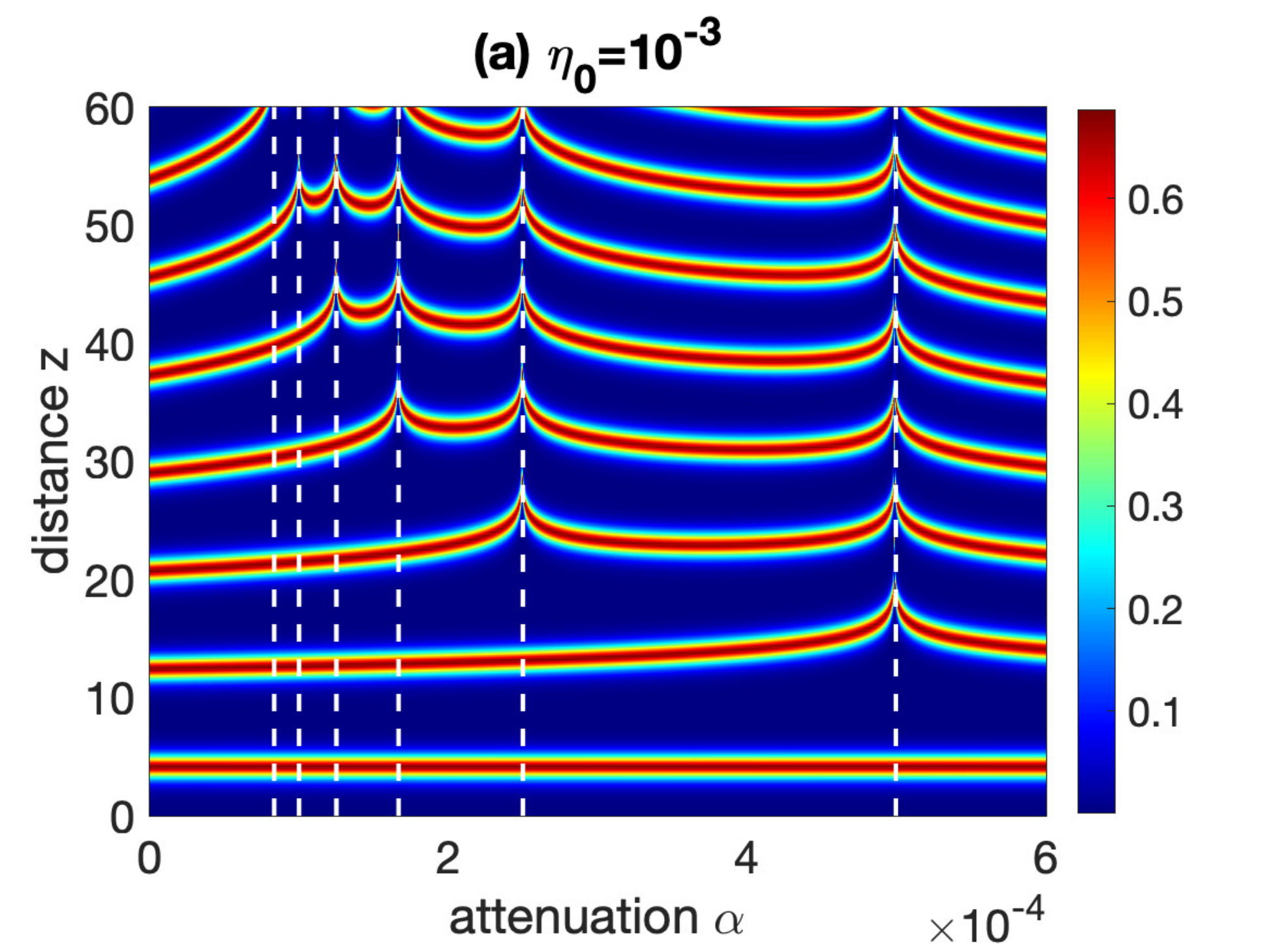}
\includegraphics[width=8.6cm]{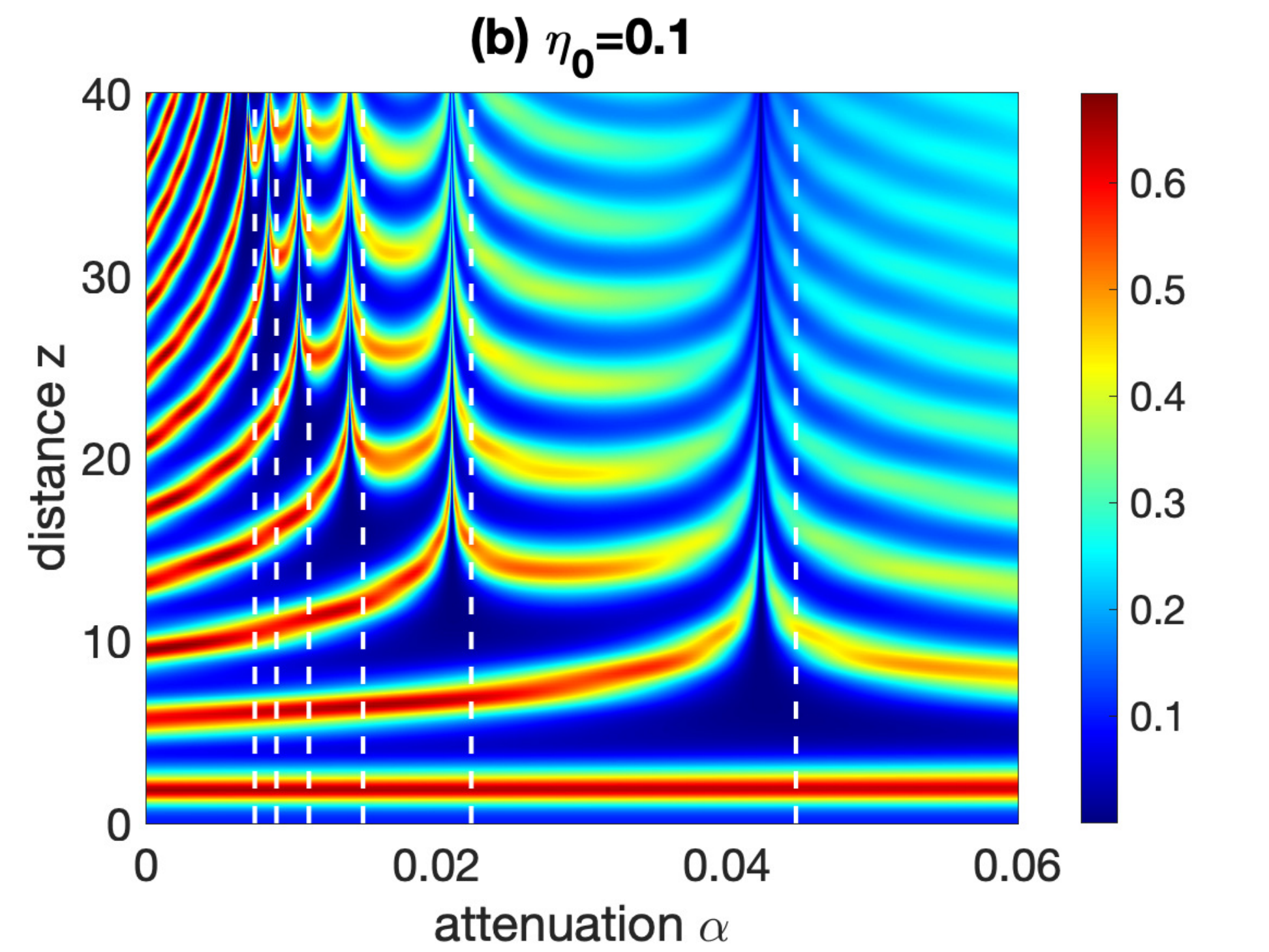}
\caption{False color plot of sideband power fraction $\eta(z)$ vs. $\alpha$ arising from numerical integration of the NLSE. 
The critical values of attenuation $\alpha_{cn}$, which correspond to the maximal distances between successive peaks of $\eta(z)$ are compared with the predictions from Eq. (\ref{alphac_IST}), reported as vertical dashed white lines. Here $\omega=\sqrt{2}$ and the initial sideband power fraction is (a) $\eta_0=10^{-3} (top)$; (b) $\eta_0=0.1 (bottom)$.}
\label{finitegap}
\end{figure}
In Eqs. (\ref{GSloss1}-\ref{GSloss3}), $z_0$ stands for the distance of the first peak amplification (i.e., with $m=1$, $z_1 \equiv z_0$), which coincides with the unperturbed expression obtained with $\alpha=0$ \cite{GS2018a,Trillo2019}, due to the assumption of small attenuation (in other words, for small $\alpha$, damping has negligible effect over the short distance $z_0$). Conversely, in Eq. (\ref{GSloss1}), the $n$-th term of the sum represents the distance $\Delta z_n$ between the $(n+1)$-th and the $n$-th peak amplifications that emerge upon evolution. In the unperturbed limit $\alpha \rightarrow 0$, $\Delta z_n$ is constant between successive peaks and represents the period of the FPUT recurrence, and depends logarithmically on the sidebands through the term $\varepsilon^2 |f_n|=\varepsilon^2 |e_+e_-|$, consistently with Refs. \cite{GS2018a,Pierangeli2018,Naveau2019b,Trillo2019}. Conversely, when $\alpha \neq 0$, due to the additional loss-dependent term in the RHS of Eq. (\ref{GSloss3}), $\Delta z_n$ is no longer constant, 
but rather becomes a non-monotone function of $\alpha$. As a result, the distances for peak amplification after the first, $z_m$ ($m=2,3,4,\ldots$), as obtained  from Eqs. (\ref{GSloss1}-\ref{GSloss3}), show a marked dependence on the damping coefficient, in qualitative agreement with Fig.~\ref{figS1}.

In the framework of this approach, the critical values of attenuation $\alpha=\alpha_{cn}$ are given by the zeros of $f_n(\alpha)$, and hence read explicitly as
\begin{equation} \label{alphac_IST}
\alpha_{cn}=  \frac{p^2 e_+ e_-}{g}  \frac{a^2}{n}, \;\; n=1,2,3,\ldots.
\end{equation}
Clearly, for  $\alpha=\alpha_{cn}$, $\Delta z_n \rightarrow \infty$ instead of becoming very large (compared to the unperturbed value) but finite as shown in Fig. 1 of the paper. Yet, Eq. (\ref{alphac_IST}) constitutes a remarkably accurate estimate of the critical losses $\alpha_{cn}$ for sufficiently weak sidebands, as shown in Fig. \ref{finitegap}(a), where we compare, for $\eta_0=10^{-3}$, the results obtained from numerical integration of the NLSE with the estimate from Eq. (\ref{alphac_IST}) reported as vertical dashed lines. However, as the sideband amplitude grow larger, the losses in Eq. (\ref{alphac_IST}) are found to overestimate the actual critical losses, as shown in Fig. \ref{finitegap}(b) for $\eta_0=0.1$, where the relative error turns to be around $\sim 15-19 \%$.

\section{Experimental setup}
\begin{figure}[ht]
\includegraphics[width=8.6cm]{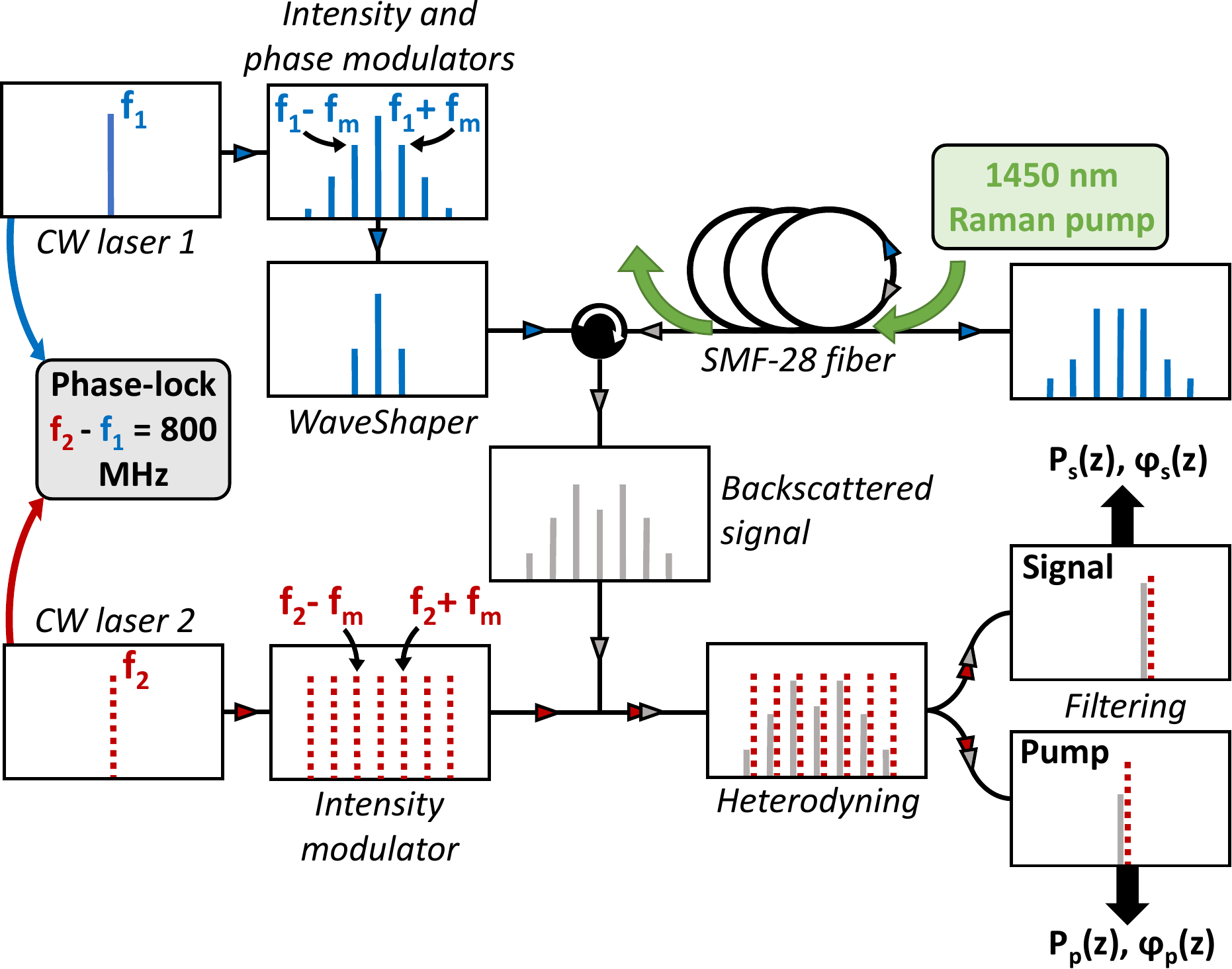}
\caption{Simplified sketch of the experimental setup. $f_{1}$ is the frequency of the main laser and $f_{2}$ of the phase-locked local oscillator, detuned by $f_2-f_1=800$ MHz. 
$f_{m}$ is the modulation frequency, set here at $38.2$ GHz. The $1450$ nm Raman pump is injected through the fiber end, and counter-propagates with respect to the main signal.}
\label{setup}
\end{figure}

This is based on a non-destructive method based on heterodyne optical time domain reflectometer (HOTDR) to resolve multiple recurrences in both power and phase along the fiber \cite{Mussot2018,Mussot2021}, on top of which we introduce an accurate control of the effective damping.
A simplified sketch of the experimental setup is presented in Fig.~\ref{setup}. 
A continuous wave laser emitting at $\lambda=1550$ nm (CW laser 1) is intensity modulated to generate a $50$ ns square pulses train (repetition rate of $4.9$ kHz), which is short enough to avoid stimulated Brillouin scattering and implement OTDR, though long enough to contain a few thousand modulation periods. The phase modulator then shapes the signal into a triangular frequency comb whose  laser line frequency spacing is $f_m$. Then, a Waveshaper, a programmable optical filter, truncates this comb to three waves only, which are tuned in amplitude and phase. Finally, an erbium doped fiber amplifier increases the total power to $528$ mW, to achieve the desired nonlinear regime (pump power $P_{P}(z=0)=480$ mW and signal power $P_{S}(z=0)=24$ mW per sideband). These light pulses then propagate into a $20.15$ km long SMF-28 optical fiber with nonlinear coefficient $\gamma=1.3$ W$^{-1}$km$^{-1}$ and group velocity dispersion $\beta_{2}=-19$ ps$^{2}$km$^{-1}$. The pump to signal frequency shift $f_m$ is set to $38.2$ GHz, close to the perfect phase matching frequency located at $41$ GHz \cite{Agrawalbook}. 

HOTDR is performed by exploiting a local oscillator (CW laser 2), phase-locked with CW laser 1, and modulated to get a frequency comb with the same line-to-line spacing $f_m$ but detuned by $800$ MHz. The power and phase distributions are obtained by heterodyning between the Rayleigh backscattered waves and the multi-tone local oscillator (see Ref. \cite{Mussot2018,Mussot2021} for details). In particular, each backscattered frequency component beats with its own nearest local oscilator spectral line. Two detection channels allow for the simultaneous recording of power and relative phase evolutions of two waves (pump and signal). 

\section{Additional experimental data}
\begin{figure}[h]
\includegraphics[width=8.6cm]{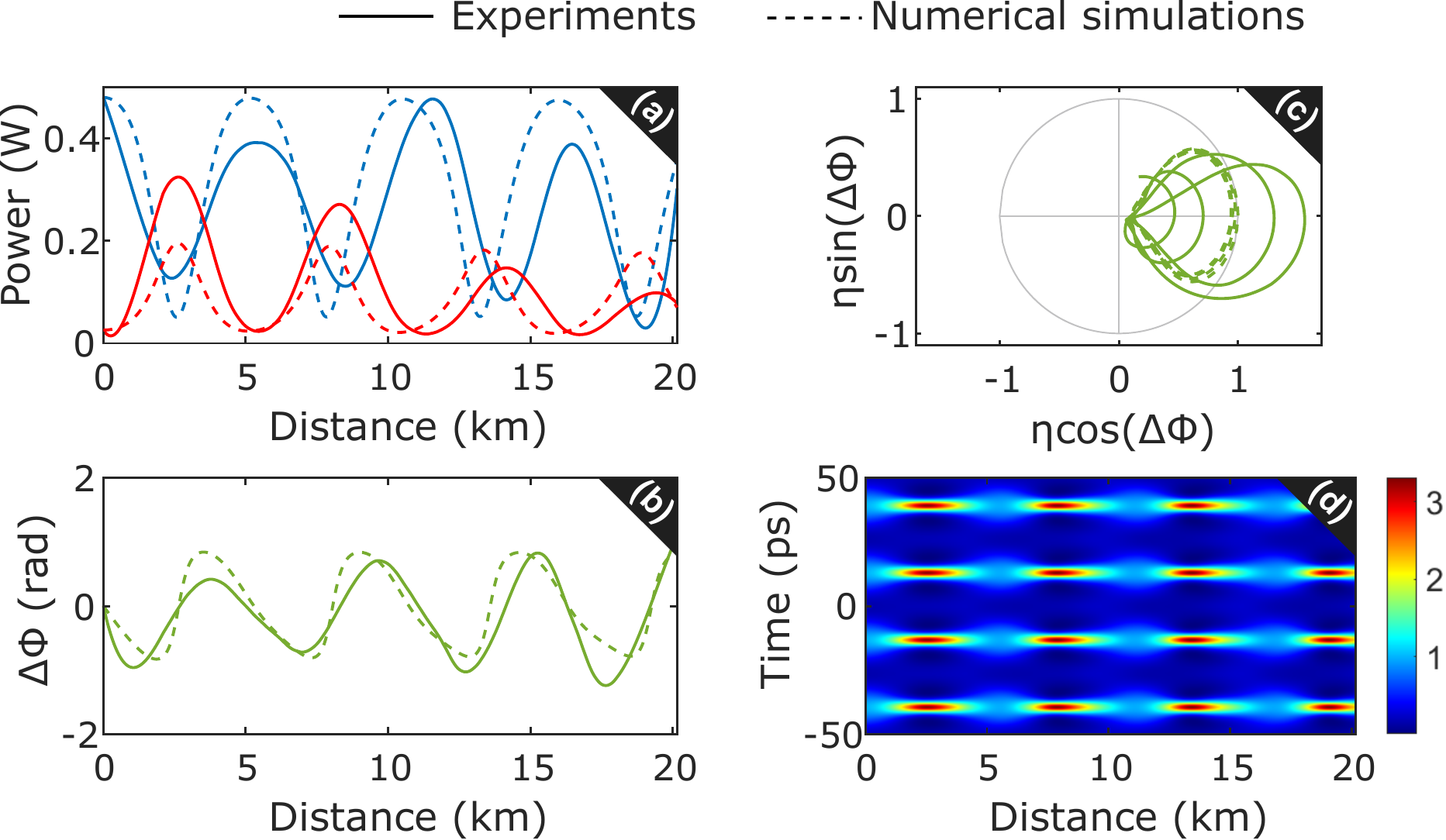}
\caption{Evolution, along the fiber length, of (a) the pump (blue lines) and the signal (red lines) powers, (b) the relative phase. (c) Phase-plane representation. (d) Numerical spatio-temporal evolution of the power. Solid lines correspond to experimental data, obtained with $P_R=280$ mW, while dashed lines to numerical simulations, with $\alpha_{eff}=0.007$ dB/km (region 3).}
\label{drawing_region3}
\end{figure}

\begin{figure}[h]
\includegraphics[width=8.6cm]{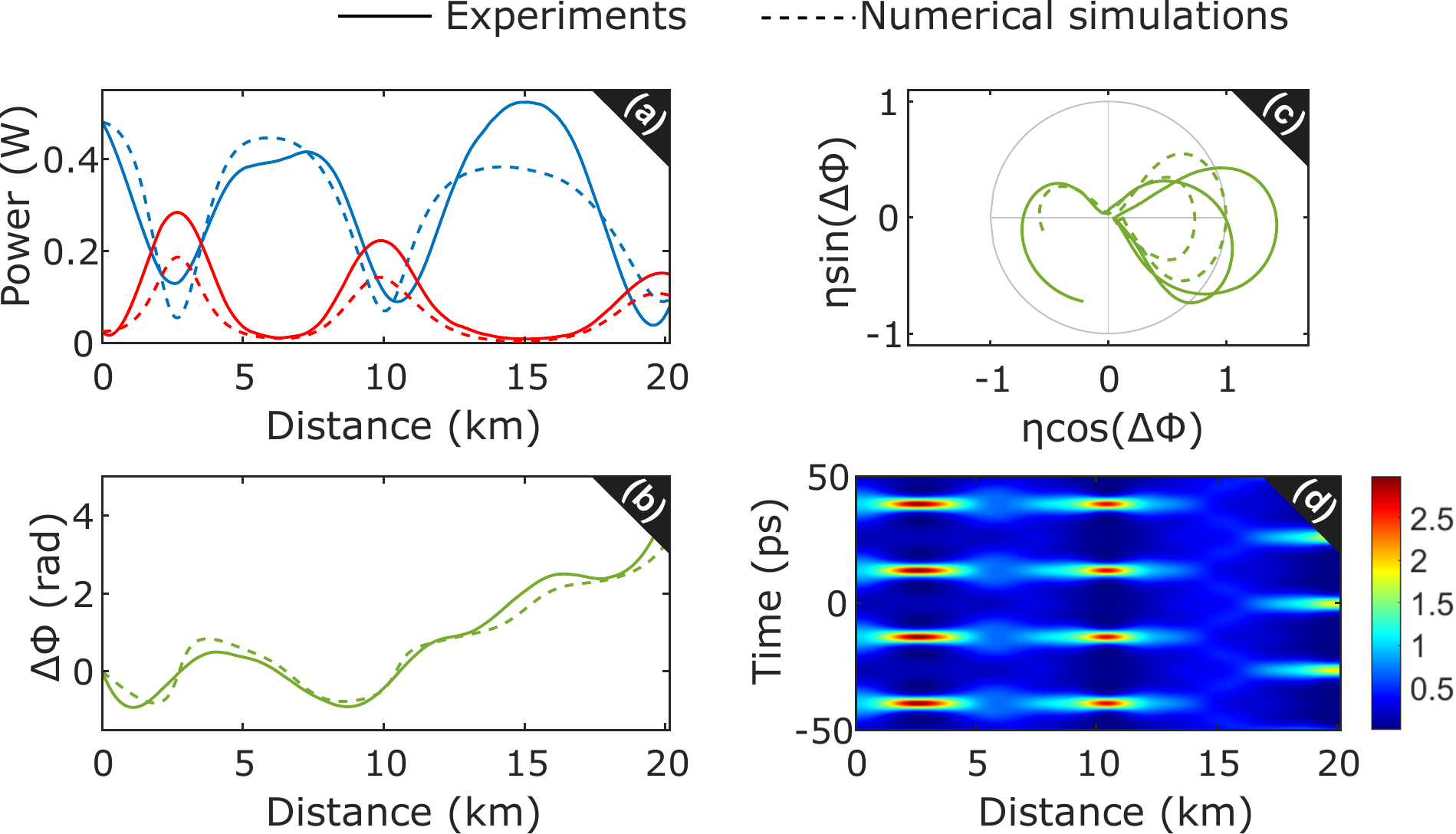}
\caption{Same as in Fig.~\ref{drawing_region3}. Experimental data are obtained with $P_R=170$ mW, numerical simulations with $\alpha_{eff}=0.09$ dB/km (region 2).}
\label{drawing_region2}
\end{figure}

\begin{figure}[h]
\includegraphics[width=8.6cm]{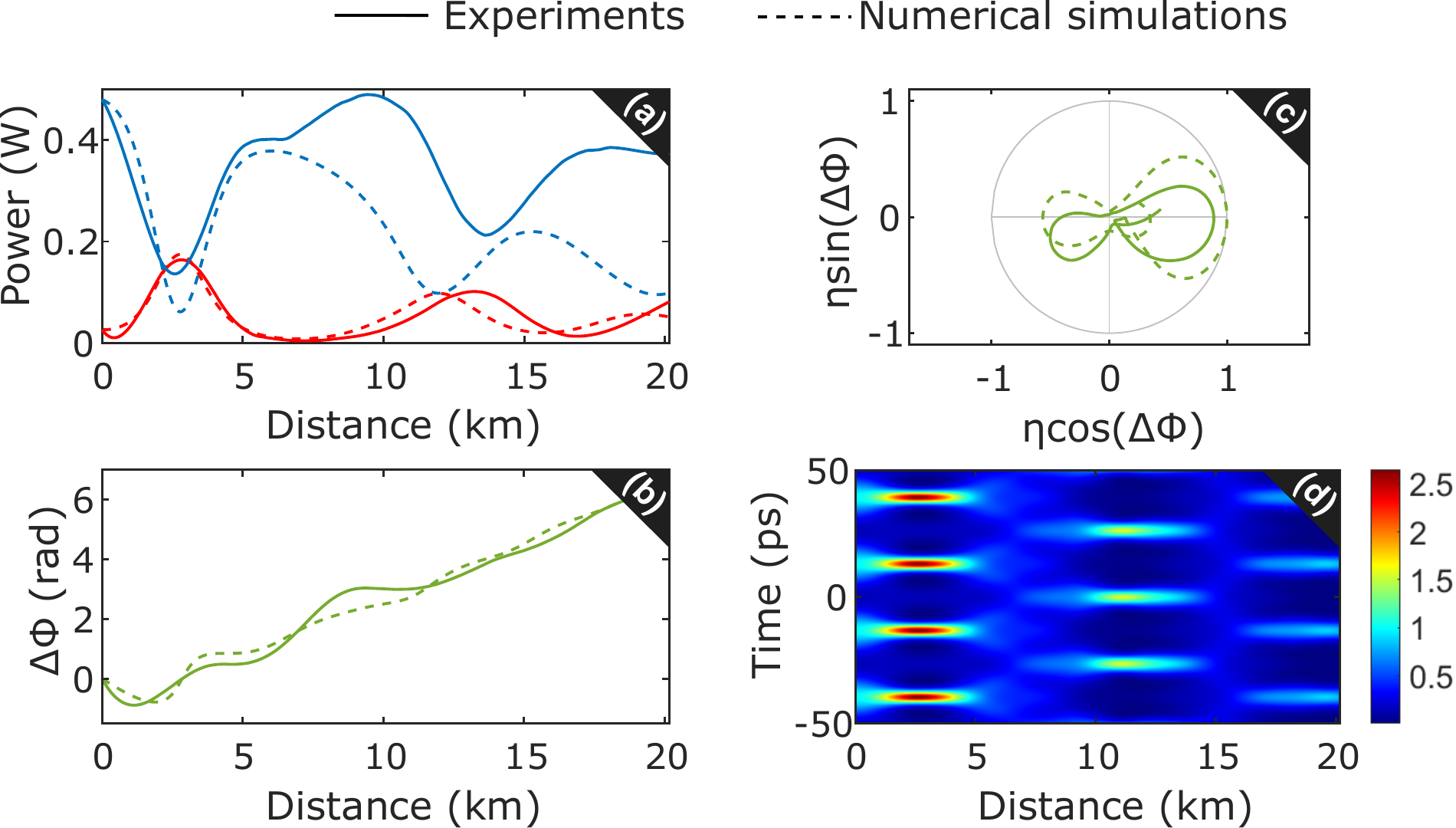}
\caption{Same as in Fig.~\ref{drawing_region3}. Experimental data are obtained with $P_R=20$ mW, numerical simulations with $\alpha_{eff}=0.2$ dB/km (region 1).}
\label{drawing_region1}
\end{figure}

Here we discuss the details of the power and phase evolutions (both measured in the experiment and numerically simulated from the NLSE), for three distinct loss values sampled in the three main regions in Fig. 4 of the paper. The results are reported in Figs.~\ref{drawing_region3}-\ref{drawing_region2}-\ref{drawing_region1}. In particular, the results presented in Fig.~\ref{drawing_region3} are performed with a quasi-perfect compensation of the losses, with $\alpha_{eff}=0.007$ dB/km (region 3).
The pump and signal power evolutions (blue and red solid lines, respectively) exhibit four sideband peak appearance (slightly less than four recurrences). Importantly, in all the FPUT periods, the experimental relative phase remains bounded (solid green line) in the range $[-\frac{\pi}{2};\frac{\pi}{2}]$ thus yielding a phase-plane trajectory confined in the right semi-plane. This means that all the recurrences are in phase (the high power pulses appears each time at the same location), as also illustrated by the numerical spatio-temporal evolution of the power in Fig.~\ref{drawing_region3}(d). 

The results presented in Fig.~\ref{drawing_region2} are relative to $\alpha_{eff}=0.09$ dB/km in region 2, obtained with $P_R=170$ mW.
In this case after two peak appearances for which the phase shows bounded oscillations, the third peak appearance is characterized by a phase that increases beyond $\pi/2$, reaches $\pi$, and continues to increase monotonically. This denotes that the separatrix has been crossed leading to a shifted orbit (when the phase is $\pi$), as also seen from the third high power pulses train $\pi$-shift in the spatio-temporal evolution of the power.

Finally, the results presented in Fig.~\ref{drawing_region1} show the case of a relatively high effective losses, namely $\alpha_{eff}=0.2$ dB/km in region 1. In this case, separatrix crossing occurs right after the first FPUT cycle, since the second peak appearance is already phase-shifted.

\section{Nonlinear MI as an inverted damped pendulum}
\hspace{-1cm}
\begin{figure}[t]
\includegraphics[width=9cm]{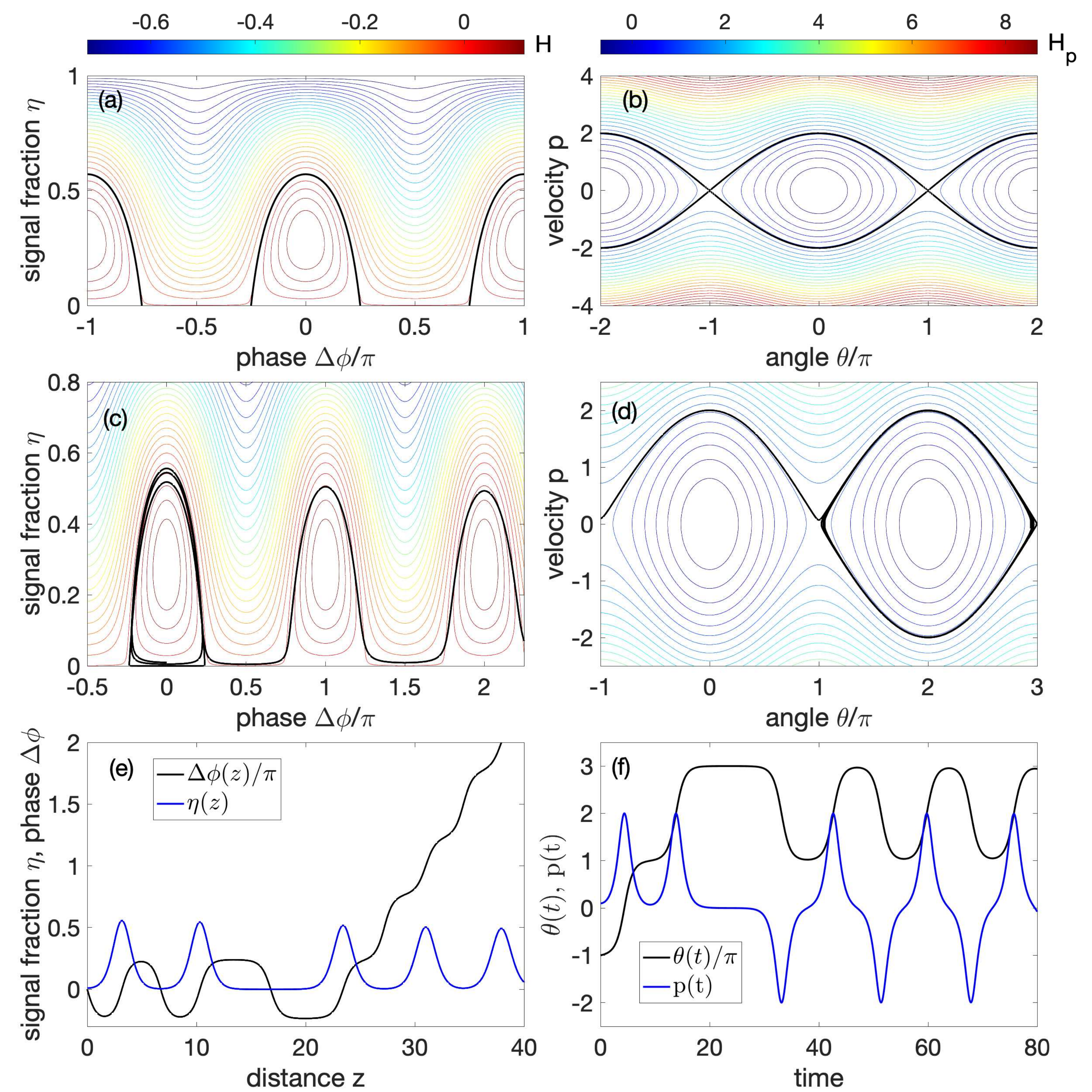}
\caption{Comparison between the damped 3WM (left panels a,c,e) and a damped pendulum (right panels b,d,f):
(a,b) Unperturbed phase-plane pictures with black solid line standing for the separatrix ((a) $\omega_0=\sqrt{2}$; (b) $\omega_0=1$); (c,d) Projection of the perturbed (by small damping) trajectories (black solid lines) on the relative phase-plane (level curves are unperturbed); (e,f) Corresponding separatrix-crossing evolutions in terms of variables (e) $\eta(z),\phi(z)$ for MI, and (f) $\theta, \dot{\theta}$ for the pendulum. Plot of MI trajectories in (c,e) are obtained with $\omega_0=\sqrt{2}$, $\alpha=4~10^{-3}$, and initially bounded or unshifted orbit (initial condition $\eta_0=0.01, \Delta \phi_0=0$), while for the pendulum $\alpha=6.2~10^{-4}$, $\omega_0=1$ and the orbit is initially unbounded (rotating pendulum, $\theta_0=-\pi$, $p_0=\dot{\theta}_0=0.1$).}
\label{cfpendulum}
\end{figure}

In spite of the infinite-dimensional nature of the problem, nonlinear MI behaves essentially like the most common of the nonlinear oscillators, namely the simple pendulum, though with opposite reaction to the effect of damping.
To show this, let us start from the motion of a standard rigid pendulum in the presence damping, which is ruled by the following Newtonian equation for the angle $\theta=\theta(t)$ (with respect to the stable equilibrium position) \cite{Butikov1999}
\begin{equation} \label{pendulum}
\ddot{\theta} + \frac{\alpha}{2} \dot{\theta} + \omega_0^2 \sin \theta = 0,
\end{equation}
where $\omega_0=\sqrt{mgh/J}$ is the frequency of the undamped small oscillations, $J$ is the moment of inertia
($J=mh^2$ for a simple pendulum of mass $m$ held at distance $h$ from centre of rotation), and $\alpha$ is the normalized energy damping rate due to friction. When $\alpha=0$, Eq. (\ref{pendulum}) gives a Hamiltonian system for conjugated variables $\theta(t)$ and angular velocity $p(t)\equiv \dot{\theta}(t)$, with Hamiltonian (conserved energy) $H_p=p^2/2-\omega_0^2 \cos \theta$. The level curves of $H_p$ in the plane ($\theta, p$), reported in Fig. \ref{cfpendulum}(b), give the classical phase-plane portrait of the pendulum. We compare such portrait with the phase-plane picture of nonlinear MI shown in Fig. \ref{cfpendulum}(a), which follows from the truncated Hamiltonian $H$ in Eq. (\ref{H}) with $\alpha=0$ (i.e., constant $\omega=\omega_0$). In order to make the comparison easier, we report the latter in the plane ($\phi,\eta$), where the saddle points in the origin in Fig. 1(a) maps into the saddles $\Delta \phi=\pm \pi/4$, which represent the stable and unstable manifolds of the pump eigenmode $\eta_s=0$ (at peak gain frequency $\omega_0=\sqrt{2}$).
The comparison of Fig. \ref{cfpendulum}(a) and (b) clearly show that, in both cases, inner domains (with respect to the separatrices, solid black lines) exist, which describe bounded motion (in angle $\theta$ for the pendulum, or phase $\Delta \phi$ for MI). They are separated from outer domains where $\theta$ or $\Delta \phi$ are instead unbounded or free-running. The inner orbits correspond to unshifted evolution of MI and librations of the pendulum, whereas the outer orbits give shifted evolution of MI and pendulum rotations. Notwithstanding the similarity between librations and unshifted orbits on one hand, and rotations and shifted orbits on the other hand, we emphasize, however, that the energetic behavior is opposite for MI and the pendulum. Indeed, as colorbars in Fig. \ref{cfpendulum}(a,b) indicate, rotations are more energetic (or more nonlinear, so to say) than librations, whereas unshifted orbits are more energetic (nonlinear) than the shifted ones. In turn, this also determines an opposite reaction to the damping. Indeed the losses always induce the energy (Hamiltonian) to decrease. As a consequence, in the pendulum, damping can only induce initial rotations to crossover into librations, but not viceversa. An example where the pendulum switches to librations after two complete rotations is shown in Figs. \ref{cfpendulum}(d,f). In particular, Fig. \ref{cfpendulum}(d) displays the phase-plane projection of the motion, whereas Fig. \ref{cfpendulum}(f) shows the temporal evolutions of angle and velocity. Conversely, nonlinear MI undergoes damping-induced separatrix-crossing from bounded (unshifted) phase evolutions to unbounded (shifted) ones [see example in Fig. \ref{cfpendulum}(c,e)].
In particular, the evolution of $\Delta \phi(z)$ in Fig. \ref{cfpendulum}(c,e) clearly marks the transition from bounded to unbounded motion, in contrast with the evolution of $\theta(t)$ in Fig. \ref{cfpendulum}(f) that exhibits the opposite transition. Therefore, we conclude that FPUT in nonlinear MI in the presence of damping exhibits a reversed behavior compared to a damped pendulum.

Finally, we remark that, even for a simple pendulum, the fine tuning of the damping is challenging, and we are not aware of experimental results that demonstrate a controlled separatrix crossing for such system.

\bibliography{dampedFPUT_PRL}
\bibliographystyle{apsrev4-1}